\documentclass[12pt]{article}
\usepackage[dvips]{graphicx}
\usepackage{amsmath}
\usepackage{graphicx}
\usepackage{amsfonts}
\usepackage{amssymb}
\usepackage{epsfig}
\usepackage{subfigure}

\begin{document}

\title{\bf Comments on holographic star and the dual QGP}
\author{Piyabut Burikham$^{1}$\thanks{Email:piyabut@gmail.com}\hspace{1mm} and Tossaporn Chullaphan$^{1,2}$\thanks{Email:chullaphan.t@gmail.com} \\
$^1$ {\small {\em  Department of Physics, Faculty of Science}}\\
{\small {\em Chulalongkorn University, Bangkok 10330, Thailand}}\\
$^2$ {\small {\em  Institute for Fundamental Study, Naresuan University, Phitsanulok, Thailand}}\\} \maketitle

\begin{abstract}
We study static AdS star in generic dimension.  The dependence of the mass limit to the bulk fermion mass is explored.  In the bulk conformal limit, the mass limit saturates at a value identical to the mass limit of a radiation star or the AdS space filling with pure radiation.  The temperature and entropy of the degenerate AdS star in the bulk conformal limit is zero in contrast to the radiation star.  Holographically, the universal mass limit corresponds to the upper limit of the deconfinement temperature in the dual gauge picture.  The QGP at this temperature is dual to the large black hole and the heat capacity is positive.  When the fermion mass increases, the mass limit falls into the range of the small black holes.  We found that even though the small black hole has negative heat capacity, the AdS box allows possibilities that it remains in thermal equilibrium with the radiation as long as the size of the black hole is not smaller than a critical size.  Consequently, the dual QGP with negative heat capacity can be produced and remains stable thermodynamically at temperature below a saturation temperature $T_{2}$.  The QGP with negative heat capacity produced at higher temperature will still condensate completely into a gas of confined hadron.      \vspace{5mm}


\newpage

\end{abstract}

\section{Introduction}

Holographic duality is one of the most exciting and rapidly advancing area of theoretical physics in recent years.  The idea of holographic duality was originated when Bekenstein argued that a black hole should possess entropy proportional to its horizon area instead of volume of the black hole~\cite{Bekenstein:1973ur}.  Intriguingly, the entropy associated with the horizon is measured in the unit of the Planck area, $l_{P}^{2}$, signalling the quantum gravity effect even for a macroscopically large black hole.  It was proposed by 't Hooft~\cite{'tHooft:1993gx} that the quantum gravity effect is revealed by the reduction of dimension occurring at the horizon.  Within a region of space, a finite amount of energy can be filled before it undergo a gravitational collapse.  Since the entropy of a black hole is determined by the horizon area, the gravitational collapse gives the upper bound on the amount of entropy or information we can store in a region of space.  It is given by the entropy of the black hole filling up the entire region, proportional to the horizon area in the unit of Planck area.  In this sense, all of the information in a region of space can be encoded in the boundary of the region, a holographic relation~\cite{Susskind:1994vu}.   

View from the observer hovering in the spacetime outside a black hole, the horizon is an effective boundary of the space.  Namely, any object falling to the black hole will appear ever more frozen and smeared as it approaches the horizon and nothing will pass through.  The holographic relation between the bulk space and its boundary can be naturally generalized to the other cases where there is no black hole in a bounded space.  One notable situation is the AdS space.  The gravity of the background AdS space pulls everything back to the center.  The massless particles can reach the infinity but they will go back to the original point in finite time.  The AdS space thus acts like a confining box and the infinity can be thought of as the boundary of the space.  In context of the string theory, the evidences of the duality between gravitational physics in the bulk and the gauge theory on the boundary of the AdS space were found by many authors~\cite{Gubser:1996de,Klebanov:1997kc,Gubser:1997yh,Gubser:1997se} which ultimately lead to the AdS/CFT correspondence conjecture by Maldacena~\cite{Maldacena:1997re,Witten:1998qj}.

An extension to finite temperature duality can be achieved by performing the path integral of the partition function in the Euclideanized time~\cite{Witten:1998zw}.  If the AdS/CFT correspondence is correct, the partition function of the gravity theory in the AdS will be the same as the partition function of the gauge theory on the boundary.  By identifying the regulated gravity action with the free energy~(modulo a $\beta$ factor) of the thermal gravity system~\cite{Hawking:1982dh}, we can study thermodynamics of the dual gauge theory on the boundary.  It is known that AdS space has 3 thermal phases; pure radiation, small black hole~(SBH), and large black hole~(LBH).  On the boundary, the bulk thermal phases of radiation and black holes correspond to the confined and deconfined phases of the strongly interacting gauge matter.  Holographically, radiation in the thermal AdS corresponds to the gas of confined hadrons whilst the radiation in the AdS-BH background corresponds to the gas or soup~(due to the strong coupling) of deconfined quarks and gluons which henceforth we will simply refer to as the QGP.  The Hawking-Page transition of the pure radiation to the black hole in the bulk AdS space is dual to the deconfinement phase transition of the confined gauge matter to the quark-gluon plasma~(QGP) on the boundary.   

The transition from the pure radiation to BHs can occur both classically by gravitational collapse and quantum mechanically by tunneling.  Conventional viewpoint when the back-reaction of the Hawking radiation is neglected is the following.  When the temperature is sufficiently large, i.e. above the critical temperature~($T_{crit}$ given by Eqn.~(\ref{Tcrit}) in this article), the pure radiation inevitably collapses to form a BH.  If the temperature is lower than $T_{crit}$ but larger than a minimum temperature $T_{\rm min}$~(given by Eqn.~(\ref{tmin}) in our text), a tunneling from pure radiation to BH could occur with a probability.      

Classical transition by gravitational collapse is a general phenomena not limited to the pure radiation.  We can consider AdS star with arbitrary particle content, fermions and bosons, and study its collapse.  Holographic duality should apply to any of these AdS stars.  A pioneering work on the holographic degenerate star with fermion content has been done in Ref.~\cite{deBoer:2009wk,Arsiwalla:2010bt} where the authors successfully construct multitrace composite operators in the CFT side as the dual of the bulk fermionic star.  AdS/CFT dictionary leads to the exact matching of the conformal dimension of the composite operator and the ADM mass of the AdS star when self-gravity is neglected.  

Holographically, gravitational collapse of an AdS star is dual to the thermalization process of the gauge matter on the boundary~\cite{dkk,ssz,ls,cy,bm,bbb,bbbc}.  The mass limit of the AdS star corresponds to the minimal amount of energy density required for the dual gauge matter to initiate the thermalization process.  Additionally, the temperature of the resulting QGP is also determined by the mass limit of the pre-collapsed star.  Investigation on the conditions of the AdS star before the gravitational collapse could arguably reveal the necessary conditions for the formation of the deconfined QGP in the dual picture.  

In this article, we extend the investigation of the fermionic AdS star to the bulk conformal limit where the bulk fermion mass is vanishing.  By numerically solving the Tolman-Oppenheimer-Volkoff~(TOV) equation, we study the effects of various parameters to the mass limit of the degenerate AdS star.  Notably by varying the fermion mass, the corresponding mass limit ranges from LBH to SBH.  In the bulk conformal limit, the mass limit saturates to a maximum value specific to each dimension.  We demonstrate that the maximal mass limit is exactly the same as the mass limit of the pure radiation as a result of the bulk conformal symmetry.  In contrast to the radiation star, the degenerate star of massless fermion has zero bulk temperature and entropy.  We interpret the maximal mass limit as the {\it universal} conformal dimension of the composite operators in the CFT side above which deconfinement phase transition is inevitable.  A remarkable linear relationship between the mass limit and total particle number is also found in the bulk conformal limit.  Bulk and boundary explanations are given.  

The article is organized as the following.  In Section \ref{II}, we briefly review the TOV equation and the equations of state of the degenerate star in the AdS space.  For completeness, we also review the construction of the composite multitrace operators dual to the degenerate star.  Rigorous calculations of the mass of the AdS star neglecting self-gravity are demonstrated to be exactly the same as given via the AdS/CFT dictionary.  In Section \ref{adsself}, we present the main results of this article.  Numerical results are interpreted in terms of the CFT on the gauge theory side.  Section \ref{IV} reviews the thermodynamics of the AdS space in generic $d$ dimensions, Eqn.~(\ref{IBH}) is a new result suitable for calculating tunneling rate of the quantum mechanical phase transition.  The bulk conformal limit is discussed in Section \ref{V}.  We also argue the possibility of stable SBH phase when the back reaction of the Hawking radiation is included.  Section \ref{VI} concludes our article.  

\section{Holographic Degenerate Star in the AdS space}  \label{II}

In this section, we consider degenerate star in the AdS space.  Since we are interested in the construction of the gravity dual of the charge-neutral gauge matter living on the boundary, we will assume that the star content is purely fermionic with no charges.  The fermion has no other interactions except gravity, both from the background AdS and the self-gravity.  The AdS star has been studied extensively in Ref.~\cite{Arsiwalla:2010bt} for zero temperature and in Ref.~\cite{Burikham:2012kn} for finite temperature with external magnetic field.  In this section, we will review the basic of the AdS star at zero temperature and discuss the situation when self-gravity is negligible.  In subsequent sections, we extend the investigation to include self-gravity and explore the limit when the fermion has zero mass.  

We also review the construction of the composite operator in the dual CFT side as suggested by Ref.~\cite{Arsiwalla:2010bt} and demonstrate in details the exact agreement between the conformal dimension of the composite operator and mass of the AdS star when self-gravity is neglected.  The review content will be extended to describe the duality in the zero-fermion-mass limit in subsequent sections, especially in the double scaling limit of the CFT side.

\subsection{The Equations of Hydrostatic Equilibrium for a Spherical Symmetric Star in d dimensions}

The equations of motion of our system are derived from the Einstein’s field equation in $d$-dimensional AdS space.  Starting from a generic spherically symmetric line element in $d$-dimension
    \begin{eqnarray}
        ds^{2} = A^{2}(r)dt^{2} - B^{2}(r)dr^{2} - r^{2}d\theta_{1}^{2} - r^{2}\sin^{2}\theta_{1}\left(d\theta_{2}^{2} + \sum_{j = 3}^{d - 2} \prod_{i = 2}^{j - 1}\sin^{2}\theta_{i}d\theta_{j}^{2}\right),
    \end{eqnarray}
We assume the energy-momentum tensor of the fermion is given by the perfect fluid form,
    \begin{eqnarray}
        T_{\mu\nu} = \left(\rho + P\right)u_{\mu}u_{\nu} + Pg_{\mu\nu},
    \end{eqnarray}
where $u_{\mu}$, $\rho$, and $P$ are the 4-velocity, energy density, and pressure respectively.  Solving the Einstein's field equation~(see e.g. Ref.~\cite{Burikham:2012kn}), we obtain the Tolman-Oppenheimer-Volkoff equation~(the TOV equation)
    \begin{eqnarray}
        \frac{dP}{dr} + \frac{1}{A}\frac{dA}{dr}\left(\rho + P\right) = 0,
    \end{eqnarray}
and the coupled equations between mass and chemical potential
    \begin{subequations}
    \begin{align}
        M'\left(r\right) &= V_{d-2}\rho\left(r\right)r^{d-2}, \label{EoM of M} \\ \mu'\left(r\right) &= \mu\left(r\right)\left(\frac{B'(r)}{B(r)} - \frac{V_{d-2}C_{d-1}}{2}\left(\rho\left(r\right)c^{2} + P_{r}\left(r\right)\right)rB^{2}\left(r\right)\right), \label{EoM of mu}
    \end{align}
    \end{subequations}
where
    \begin{subequations}
    \begin{align}
        B(r) &= \left(1 - \frac{M(r)C_{d - 1}}{r^{d - 3}} + \frac{r^{2}}{\ell^{2}}\right)^{- 1/2}, \\
        A^{2}(r) &= \frac{e^{2\chi(r)}}{B^{2}(r)}, \\
        \chi(r) &= \frac{V_{d - 2}C_{d - 1}}{2}\int\left(\rho + P\right)rB^{2}(r)dr.
    \end{align} \label{bmetric}
    \end{subequations}
The parameter $\ell$, $V_{d - 2}$, $C_{d - 1}=\frac{16\pi G}{(d-2)V_{d-2}c^{4}}$~\footnote{the gravitational constant $G$ here is chosen to be $2/(d-2)$ of the constant used in Ref~\cite{Burikham:2012kn} so that it is in the same convention as the one in Eqn.~(\ref{gact})} and $M(r)$ are the radius of the AdS space, the area of $S^{d - 2}$, a constant and the accumulated mass of the star respectively.  The chemical potential is naturally redshifted by the metric as 
    \begin{eqnarray}
        \mu = \frac{\epsilon_{F}}{A(r)}, \label{mu and A}
    \end{eqnarray}
where we identify the central chemical potential with the Fermi energy $\epsilon_{F}$ for empty AdS with $A(0)=1$.

For a fermionic star, the density and pressure of the fermion gas at zero temperature are given by 
\begin{eqnarray}
\rho & = & \frac{g_{f}V_{d-2}}{(2\pi)^{d-1}}\int_{m}^{\mu}~\mu^{2}(\mu^{2}-m^{2})^{(d-3)/2}~d\mu,  \label{frho}  \\
P & = & \frac{g_{f}V_{d-2}}{(d-1)(2\pi)^{d-1}}\int_{m}^{\mu}~(\mu^{2}-m^{2})^{(d-1)/2}~d\mu,  \label{fp}
\end{eqnarray}
where $g_{f}$ is the number of internal degrees of freedom such as spin, flavour and colour of the fermion.  Note that for $m=0$~(conformal limit), the density and pressure become
\begin{eqnarray}
\rho = \frac{g_{f}V_{d-2}}{(2\pi)^{d-1}}\frac{\mu^{d}}{d}, P = \frac{\rho}{d-1},  \label{zerom}
\end{eqnarray}
a linear equation of state.  In $m=0$ limit, the AdS star behaves like a radiation star~(at zero temperature, this is possible for a fermionic star or a Bose-condensate star).  The radiation star in the AdS space turns out to have a finite mass limit but infinite radius as we will see later.  Because of the infinite radius, the radiation star can be thought of as a thermal phase of the AdS space, i.e. the AdS space filling with radiation.  The temperature profile of the radiation star is given by $T(r)=T(0)/A(r)$.  We will assume that $T=0$ when we consider the AdS star and only turn on the temperature when we compare the AdS-BH with the thermal AdS in subsequent section.

\subsection{Mass of degenerate star between bulk and boundary in the absence of self-gravity} \label{cor0}

According to the usual AdS/CFT dictionary, the conformal dimension of the fermionic single trace operator $\Delta_{0}$ is related to the mass of the bulk fermion $m$ by 
\begin{eqnarray} 
\Delta_{0}&=&\frac{d}{2}+\sqrt{\frac{d^{2}}{4}+(m \ell)^{2}}. \label{mdcorr}
\end{eqnarray}
For large $m\gg d/2\ell$, $\Delta_{0}\simeq m \ell$, the conformal dimension of an operator can be thought of as a scaled mass in the bulk.  

By using the dictionary above, Arsiwalla et al \cite{Arsiwalla:2010bt} found a perfect matching between the bulk mass of a spherical symmetric AdS star and the conformal dimension of the multi-trace operator in the boundary description when the self-gravity of the bulk is neglected.  Let us briefly review how they construct the composite operators and count the conformal dimensions of such operators.  Exclusion principle demands that the composite operators, $\Phi$, should contain derivatives of the single-trace fermionic operators, $\Psi$~(since the CFT exists in $(d-1)$-dimensions, the number of possible derivatives is $d-1$).  The most generic and simplest form is given by
\begin{eqnarray}
\Phi & = & \Psi \prod_{i}\partial_{i}\Psi \prod_{\{i,j\}}\partial_{i}\partial_{j}\Psi...\prod_{\{i_{1},i_{2},...,i_{n}\}}\partial_{i_{1}}\partial_{i_{2}}...\partial_{i_{n_{F}}}\Psi
\end{eqnarray}
where the product over all possible permutation of the derivatives is included for a given number of derivatives.  Each product contains different numbers of derivatives of the single-trace operator and we call them a ``shell".  The last shell contains permutation of $n$ derivatives at the most and $n_{F}$ is the maximal number of derivatives in the last shell.  The composite operator thus depends only on $n_{F}$ and the number of possible configuration for a fixed $n_{F}$ is one.  The entropy of each $n_{F}$-operator is zero so we call it a {\it degenerate} multi-trace operator.  A degenerate multi-trace operator is a good candidate for the corresponding dual of the degenerate fermion gas~(at zero temperature) in the bulk.

The number of fields $\Psi$ in the shell with $n$ derivatives is given by 
\begin{equation*}
\left( \begin{array}{c} n+d-2  \\
d-2 \end{array}
\right),
\end{equation*}
which is equal to the number of particles in the bulk.  For a fixed $n_{F}$, the total number of fields corresponding to the total number of bulk particles is therefore
\begin{eqnarray}
N_{F}& = & \sum_{n=0}^{n_{F}}\left( \begin{array}{c} n+d-2  \\
d-2 \end{array}
\right)=\left( \begin{array}{c} n_{F}+d-1  \\
d-1 \end{array}
\right). \label{NF}
\end{eqnarray}
In order to estimate the bulk mass, we need to calculate the conformal dimension of the composite operator $\Phi$ for a certain $n_{F}$.  Since each derivative increases conformal dimension by one, the conformal dimension of $\Phi$ is
\begin{eqnarray}
\Delta & = & \sum_{n=0}^{n_{F}}(n+\Delta_{0})\left( \begin{array}{c} n+d-2  \\
d-2 \end{array}
\right)=(d-1)\left( \begin{array}{c} n_{F}+d-1  \\
d \end{array}
\right)+\Delta_{0}\left( \begin{array}{c} n_{F}+d-1  \\
d-1 \end{array}
\right).
\end{eqnarray}

By using the approximate correspondence~(valid only when $m\gg d/2\ell$)
\begin{eqnarray}
m \simeq \frac{\Delta_{0}}{\ell}, M \simeq \frac{\Delta}{\ell}, \epsilon_{F} \simeq \frac{n_{F}+\Delta_{0}}{\ell},  \label{corr}
\end{eqnarray}
we find an approximate relation between the bulk mass, the Fermi energy and the bulk fermion mass to the leading order of $\ell$ as the following
    \begin{eqnarray}
        M = \frac{\left(d - 1\right)\ell^{d - 1}}{d!}\left(\epsilon_{F} - m\right)^{d} + \frac{m\ell^{d - 1}}{\left(d - 1\right)!}\left(\epsilon_{F} - m\right)^{d - 1}. \label{CFT Mass}
    \end{eqnarray}
where 
\begin{eqnarray}
\left( \begin{array}{c} n_{F}+d-1  \\
d \end{array}
\right)&\simeq& \frac{n_{F}^{d}}{d!}, \nonumber \\
\left( \begin{array}{c} n_{F}+d-1  \\
d-1 \end{array}
\right)&\simeq& \frac{n_{F}^{d-1}}{(d-1)!} \label{app}
\end{eqnarray}
are used.  The term proportional to $m$ corresponds to the rest energy and the term proportional to $d-1$, originated from derivatives, corresponds to the kinetic energy of the fermion in the bulk.      

On the other hand, the gravitational mass of the AdS star in the bulk can be calculated via Eqn.~(\ref{EoM of M}).  Integrating by parts and using thermodynamic relation $d\rho = \mu ~dn$, we have
    \begin{eqnarray}
        M = \frac{V_{d - 2}~\rho ~r^{\left(d - 1\right)}}{\left(d - 1\right)}\Big|_{0}^{R} - \int_{\rho(r = 0)}^{\rho(r = R)}\frac{V_{d - 2}~r^{\left(d - 1\right)}}{\left(d - 1\right)}d\rho
        = -\int_{\epsilon_{F}}^{m}\frac{V_{d - 2}~r^{\left(d - 1\right)}}{\left(d - 1\right)}\mu\frac{dn}{d\mu}d\mu. \nonumber
    \end{eqnarray}
The number density in $d$-dimension is given by 
\begin{eqnarray}
n = \frac{V_{d-2}}{(d - 1)(2\pi)^{(d - 1)}}\left(\mu^{2} - m^{2}\right)^{\frac{(d-1)}{2}}. \label{nden}
\end{eqnarray}
Using (\ref{mu and A}) and solving $r(\mu)$ when ignoring self-gravity, i.e. $\chi(r) = 0, r = \ell\sqrt{\left(\frac{\epsilon_{F}}{\mu}\right)^{2} - 1}$, we get
    \begin{eqnarray}
        M = -\frac{\left(V_{d - 2}\right)^{2}\ell^{d - 1}}{\left(d - 1\right)\left(2\pi\right)^{\left(d - 1\right)}}\int_{\epsilon_{F}}^{m}\left(\left(\frac{\epsilon_{F}}{\mu}\right)^{2} - 1\right)^{\frac{d - 1}{2}}\mu^{2}\left(\mu^{2} - m^{2}\right)^{\frac{d - 3}{2}}d\mu. \label{mbu}
    \end{eqnarray}
After performing the integration,
    \begin{eqnarray}
        M = \frac{\left(V_{d - 2}\right)^{2}m^{d}\ell^{d - 1}}{\left(d - 1\right)\left(2\pi\right)^{\left(d - 1\right)}}\frac{\sqrt{\pi}\left(\frac{\epsilon_{F}}{m} - 1\right)^{d - 1}}{4\Gamma\left(\frac{d}{2} + 1\right)}\left(1 + (d - 1)\frac{\epsilon_{F}}{m}\right)\Gamma\left(\frac{d - 1}{2}\right),
    \end{eqnarray}
where $\Gamma\left(z\right)$ is a gamma function.  By using properties of Gamma function
\begin{eqnarray}
z~\Gamma(z)=\Gamma(z+1), ~~\Gamma(z)~\Gamma\left(z+\frac{1}{2}\right)=2^{1-2z}\sqrt{\pi} ~\Gamma(2z), \label{gamid}
\end{eqnarray}
and $V_{d-2}=2\pi^{(d-1)/2}/\Gamma((d-1)/2)$, the AdS mass can be calculated to be exactly the same expression as the right-hand side of Eqn.~(\ref{CFT Mass}), i.e.
    \begin{eqnarray}
        M = \frac{\left(d - 1\right)\ell^{d - 1}}{d!}\left(\epsilon_{F} - m\right)^{d} + \frac{m\ell^{d - 1}}{\left(d - 1\right)!}\left(\epsilon_{F} - m\right)^{d - 1}.     \label{mat0}
\end{eqnarray}
This is an amazing result showing the correspondence between the conformal dimension of multitrace operator in the CFT and the mass of the AdS fermionic star in the bulk.  We should remark that the calculation of the bulk mass depends crucially on the functional form of the chemical potential with respect to the radial coordinate, $\mu(r)=\epsilon_{F}/A(r)=\epsilon_{F}/\sqrt{1+r^{2}/\ell^{2}}$ for the AdS space.  This is the main reason why there is an exact matching between the bulk mass and the mass calculated from conformal dimension via the dictionary of the AdS/CFT correspondence.  

We can rescale the total bulk mass of the AdS star in terms of the bulk fermion mass $m$ as the following
\begin{eqnarray}
M & = & \frac{1}{\ell}\frac{(m\ell)^{d}}{d!}(x_{0}-1)^{d-1}(1+(d-1)x_{0});~~x_{0}\equiv \frac{\mu_{0}}{m},
\end{eqnarray}
where the central chemical potential, $\mu_{0}$, is identified with the Fermi energy $\epsilon_{F}$.  The mass is an increasing function of $\mu_{0}$ and contains no information of the mass limit of the degenerate star.  It neglects the effect of self-gravity and is not suitable for the gravitational collapse situation where self-gravity dominates and competes with the degenerate pressure.  We will consider the situation when self-gravity is included in the next section.   

If we consider $d = 4$ and $5$ cases, we obtain the mass in the $\rm{AdS}_4$ and $\rm{AdS}_5$
    \begin{subequations}
    \begin{align}
        M_{AdS_{4}} &= \frac{l^{3}}{24}\left(\epsilon_{F} - m\right)^{3}\left(3\epsilon_{F} + m\right), \\
        M_{AdS_{5}} &= \frac{l^{4}}{120}\left(\epsilon_{F} - m\right)^{4}\left(4\epsilon_{F} + m\right),
    \end{align}
    \end{subequations}
respectively.  

The total particle number $N_{F}$ can also be integrated with respect to the curved spatial volume by using the number density given by Eqn.~(\ref{nden}),
\begin{eqnarray}
N_{F}& = & V_{d-2}\int_{0}^{R}~n(r)B(r)r^{d-2}~dr. 
\end{eqnarray}
Substitute $B(r)=\mu/\epsilon_{F}$~(since when self-gravity is neglected $AB =1$) and change the integration variable to $\mu$, we obtain
\begin{eqnarray}
N_{F}& = &  -\frac{\left(V_{d - 2}\right)^{2}\ell^{d - 1}}{\left(d - 1\right)\left(2\pi\right)^{\left(d - 1\right)}}\int_{\epsilon_{F}}^{m}\left(\left(\frac{\epsilon_{F}}{\mu}\right)^{2} - 1\right)^{\frac{d - 3}{2}}\mu^{-2}\left(\mu^{2} - m^{2}\right)^{\frac{d - 1}{2}}d\mu, \label{nbu} \\
& = & \frac{\left(V_{d - 2}\right)^{2}\ell^{d - 1}\epsilon_{F}^{d-1}}{\left(d - 1\right)\left(2\pi\right)^{\left(d - 1\right)}}\frac{\sqrt{\pi}\left(1-\frac{m}{\epsilon_{F}}\right)^{d - 1}}{2\Gamma\left(\frac{d}{2}\right)}\Gamma\left(\frac{d - 1}{2}\right),\nonumber \\
& = & \frac{\ell^{d-1}}{(d-1)!}(\epsilon_{F}-m)^{d-1}, \label{ntot}
\end{eqnarray}
where we have used the identities of Gamma function, Eqn.~(\ref{gamid}), in the last step as before.  Again this is in exact agreement with the total particle number given by Eqn.~(\ref{NF}), (\ref{app}) and (\ref{corr}) in the boundary theory.  Remarkably, we can define the density of states $g(\epsilon)\equiv \ell^{d-1}(\epsilon - m)^{d-2}/(d-2)!$ for the ``boundary free field" $\Psi$~\cite{Arsiwalla:2010bt} for which
\begin{eqnarray}
N_{F} = \int_{m}^{\epsilon_{F}}g(\epsilon)~d\epsilon, \quad M = \int_{m}^{\epsilon_{F}}\epsilon ~g(\epsilon)~d\epsilon.  \label{nmbo}
\end{eqnarray}
It is interesting to compare the integrand of Eqn.~(\ref{mbu}), (\ref{nbu}), and (\ref{nmbo}) since they are integrated within the same range of energy variable.  It is quite intriguing that different functions of integrand give exactly the same result, bulk and boundary.  The boundary description has a statistical interpretation in terms of the density of states $g(\epsilon)$ while we cannot define a density-of-states function in the bulk, i.e. the integrand in Eqn.~(\ref{nbu}) does not lead to the correct integrand of mass in Eq.~(\ref{mbu}) when multiplied by $\mu$.  Nevertheless, an obvious relationship is established between the {\it bulk} mass and particle number as we can derive from differentiating Eqn.~(\ref{nmbo}) with respect to $\epsilon_{F}$ on the {\it boundary}
\begin{eqnarray}
\frac{dM}{d\epsilon_{F}} & = & \epsilon_{F}\frac{dN_{F}}{d\epsilon_{F}}.
\end{eqnarray}
The same relationship can also be obtained by using the Einstein equations as pointed out in Ref.~\cite{Arsiwalla:2010bt}.

In fact, since the chemical potential is determined as a function of the radial coordinate $r$ as 
\begin{eqnarray}
\mu(r) & = & \frac{\epsilon_{F}}{\sqrt{1+\frac{r^{2}}{\ell^{2}}}},
\end{eqnarray}
the accumulated mass $M(r)$ can be directly integrated out to be a function of $r$,
\begin{eqnarray}
M(r) & = & \frac{\left(V_{d - 2}\right)^{2}\ell^{d - 1}\epsilon_{F}^{d}}{\left(d - 1\right)\left(2\pi\right)^{\left(d - 1\right)}}\int_{u}^{1}\left(\frac{1}{u^{2}} - 1\right)^{\frac{d - 1}{2}}u^{2}\left(u^{2} - u_{R}^{2}\right)^{\frac{d - 3}{2}}du,
\end{eqnarray}
where $u\equiv \mu(r)/\epsilon_{F}, u_{R}\equiv \mu(R)/\epsilon_{F}=m/\epsilon_{F}$.  For example, when $d=5$ the integrand becomes a polynomial and the accumulated mass is given by
\begin{eqnarray}
M(r)_{AdS_{5}} & = & \frac{\ell^{4}\epsilon_{F}^{5}}{16}\frac{(u-1)^{3}}{15 u}\left(5 u_{R}^{2}(3+u)-u(3u^{2}+9u+8)\right),
\end{eqnarray}
with $u=1/\sqrt{1+r^{2}/\ell^{2}}$.  Similarly for the particle number accumulated from $r=0$ to $r$,
\begin{eqnarray}
N_{F}(r) & = & \frac{\left(V_{d - 2}\right)^{2}\ell^{d - 1}\epsilon_{F}^{d-1}}{\left(d - 1\right)\left(2\pi\right)^{\left(d - 1\right)}}\int_{u}^{1}\left(\frac{1}{u^{2}} - 1\right)^{\frac{d - 3}{2}}u^{-2}\left(u^{2} - u_{R}^{2}\right)^{\frac{d - 1}{2}}du.
\end{eqnarray}
For $d=5$, the polynomial integral is simplified to 
\begin{eqnarray}
N_{F}(r)_{AdS_{5}}& = & \frac{\ell^{4}\epsilon_{F}^{4}}{16}\frac{(u-1)^{2}}{3u^{3}}\left(u^{3}(u+2)-6u_{R}^{2}u^{2}+u_{R}^{4}(2u+1) \right),
\end{eqnarray}
with $u(r)$ given above.

It is apparent from Eqn.~(\ref{mat0}) and (\ref{ntot}) that both the total mass and particle number of the AdS star are increasing functions of $\epsilon_{F}$ when the self-gravity is neglected~(and $\mu_{0} = \epsilon_{F}$).  Consequently, there is no mass limit in this case.  The mass and particle number continue to increase without bound as the central chemical potential grows.  Naturally, neglecting self-gravity leaves us with only the gravity generated from the AdS background metric.  The AdS star can attain any large masses which are determined from the equilibrium between the background gravity and the degenerate pressure.  Since the background gravity is always bounded~(i.e. no coordinate singularities), there is no possibility of gravitational collapse signalling the appearance of the event horizon, therefore the absence of the mass limit.

\section{AdS stars with self-gravity}  \label{adsself}

In section \ref{cor0}, it is demonstrated that the correspondence between conformal dimension of the boundary operator and the mass of semiclassical bulk field lead to an exact matching between the conformal dimension of the multitrace fermionic operator and the bulk mass of the fermionic degenerate star, Eqn.~(\ref{mat0}), in the absence of self-gravity.  In obtaining the correspondence, we have made certain approximation $m\gg d/2\ell$, $\Delta_{0}\simeq m \ell$ that the mass is large.  It is also taken that $n_{F}$ is large comparing to $d$.

In the presence of strong self-gravity, there is a possibility of coordinate singularities appearing in the metric, Eqn.~(\ref{bmetric}).  Strong gravity limits the maximal possible mass of the star when the internal pressure becomes inevitably smaller than the strength of gravity.  The result is the Chandrasekhar/TOV mass limit of a degenerate star.  Certainly, when the self-gravity is strong, the basic fermionic operator, $\Psi$, must interact strongly to one another, changing the conformal dimension of the composite multitrace operator, $\Phi$, in the process.  There has been attempt to find a new conformal dimension of the strongly coupled composite operator in order to match with the mass of the AdS star in the presence of strong self-gravity~\cite{Arsiwalla:2010bt}.  Apparently in the Newtonian limit, the self-gravity contributes a negative amount of mass, roughly $-GM^{2}/r$ which corresponds to correction to the conformal dimension $\Delta_{int.}\sim - \Delta^{2}/N^{2}$ in the 't Hooft limit since $M\sim \Delta/\ell, r\sim\ell, G\sim G_{10}/\ell^{6}\sim \ell^{2}/N^{2}$.  We also have to multiply bulk energy by $\ell$ in order to obtain the conformal dimension according to Eqn.~(\ref{corr}).  Consquently, the conformal dimension of $\Phi$ corresponding to the bulk mass of the AdS star in the presence of self-gravity becomes approximately
\begin{eqnarray}  
\Delta + \Delta_{int.} &\sim& \Delta -(\text{Const.})\frac{\Delta^{2}}{N^{2}}.  \label{delself}
\end{eqnarray}
On the boundary, the change in the conformal dimension of the composite operator $\Phi$ is due to the 4-point interaction between the constituent operator $\Psi$ which can be extracted from the OPE of the 4-point function of the operator $\Psi$.  At the leading order of $1/c$-expansion~($c:$central charge of the boundary CFT), the interaction is the exchange of the stress tensor as well as its crossing channel and thus it is {\it universal} just like gravity in the bulk.   

\subsection{large $\ell$ limit} \label{lell}

When the radius of AdS space is large, the central charge of the dual CFT, $c \sim N^{2} \sim \ell^{d-2}/G$ will be large.  If $\Delta, N_{F}, \Delta_{0}$ are fixed, the fermionic operator $\Psi$ can be approximated by a ``free field" with negligible interaction.  On the other hand, if we also scale $\Delta/c =f$ to be a finite constant not much different from $\mathcal{O}(1)$, we will have
\begin{eqnarray}
\frac{\Delta}{c}& \sim & \frac{M\ell G}{\ell^{d-2}}\sim f.
\end{eqnarray}
In the bulk from Eqn.~(\ref{bmetric}), this would correspond to the situation where self-gravity of the bulk fermion gas can no longer be neglected.  Namely,
\begin{eqnarray}
\frac{R^{2}}{\ell^{2}}\simeq \frac{GM}{R^{d-3}}\sim \mathcal{O}(1).
\end{eqnarray}
In this ``double scaling limit"~\cite{Arsiwalla:2010bt}, since $k_{F}\simeq \epsilon_{F}$ we have
\begin{eqnarray}
\frac{R}{\ell}=\frac{k_{F}}{m}\simeq \frac{n_{F}}{\Delta_{0}}\simeq g({\rm const}),
\end{eqnarray}
therefore
\begin{eqnarray*}
\Delta \sim \Delta_{0}n_{F}^{d-1}, 
\end{eqnarray*}
or
\begin{eqnarray*}
n_{F} \sim \Delta_{0}\sim c^{1/d},
\end{eqnarray*}
where Eqn.~(\ref{corr}),(\ref{CFT Mass}), and (\ref{app}) have been used.  It also implies the scaling of the total particle number, Eqn.~(\ref{NF}), with respect to the central charge
\begin{eqnarray*}  
 N_{F}\sim (\ell \epsilon_{F})^{d-1}\sim \Delta_{0}^{d-1}\sim c^{(d-1)/d}.
\end{eqnarray*}
In the bulk picture, the scaling $\Delta_{0}\sim c^{1/d}$ corresponds to the relation
\begin{eqnarray}
\frac{m\ell G^{1/d}}{\ell^{(d-2)/d}}\sim {\rm finite~const,}
\end{eqnarray}
which demands that as $\ell\to \infty$, the Newton constant must scale as $G\to ({\rm const.})/\ell^{2}$ in any dimensions.  By requiring the invariance of the bulk equations of motion Eqn.~(\ref{EoM of M}), (\ref{EoM of mu}), (\ref{bmetric}) and (\ref{frho}) under scaling $r\to r r_{0}, M\to M M_{0}, m\to m m_{0}, \rho\to \rho \rho_{0}, G\to G G_{0}$, the scales must satisfy the relations
\begin{eqnarray}
\rho_{0}=m_{0}^{d}=\frac{1}{r_{0}^{2}G_{0}}, M_{0}=\frac{r_{0}^{d-3}}{G_{0}}.
\end{eqnarray}
Choosing $r_{0}=\ell$ and applying $G\ell^{2}={\rm const.}$ result in the relation
\begin{eqnarray}
\frac{M}{\ell^{d-1}}& = & {\rm const.}
\end{eqnarray}
For a fixed $m$, this is exact regardless of $\ell$ as long as we choose $G={\rm const.}/\ell^{2}$.  For large $\ell$, we have $\Delta \sim c\sim \ell^{d}, \Delta_{0}\sim \ell, N_{F}\sim \ell^{d-1}$ on the boundary CFT.  Consequently for a given $m$, we can choose sufficiently large $\ell$ so that $M$ is larger than $m$.  Another prediction is that $M/N_{F}={\rm const.}$ since both quantities are of order $\ell^{d-1}$ and the constant $f, g$ are not sensitive to $m$~(when $m$ is small).  We will see that this is actually the case when we numerically calculate the mass limit and $N_{F}$ of the AdS star with self-gravity in the next section, as shown in Fig.~\ref{mlimn}.

\subsection{fixed $\ell$ case}

Generically, the AdS/CFT correspondence should be valid even when $\ell$ is fixed to some finite constant.  In this section we will explore this possibility by setting $\ell=1$ and argue that it could lead to the quark-gluon plasma with negative heat capacity.  Increasing $\ell$ will increase the mass limit of the fermion star for a fixed $m$ but we can always increase $m$ so that the mass limit becomes substantially smaller, even smaller than $m$ itself.  In this Section, we numerically solve the coupled equations of motion, Eqn.~(\ref{EoM of M}), (\ref{EoM of mu}), by setting $G, \ell =1$.  As long as we keep $G\ell^{2}$ fixed at a specific value of $m$, the results remain unchanged for any other values of $G$ and $\ell$.

\subsection{Relations between mass limit, fermion mass and radius}

\begin{figure}[center]
        \includegraphics[scale=0.9]{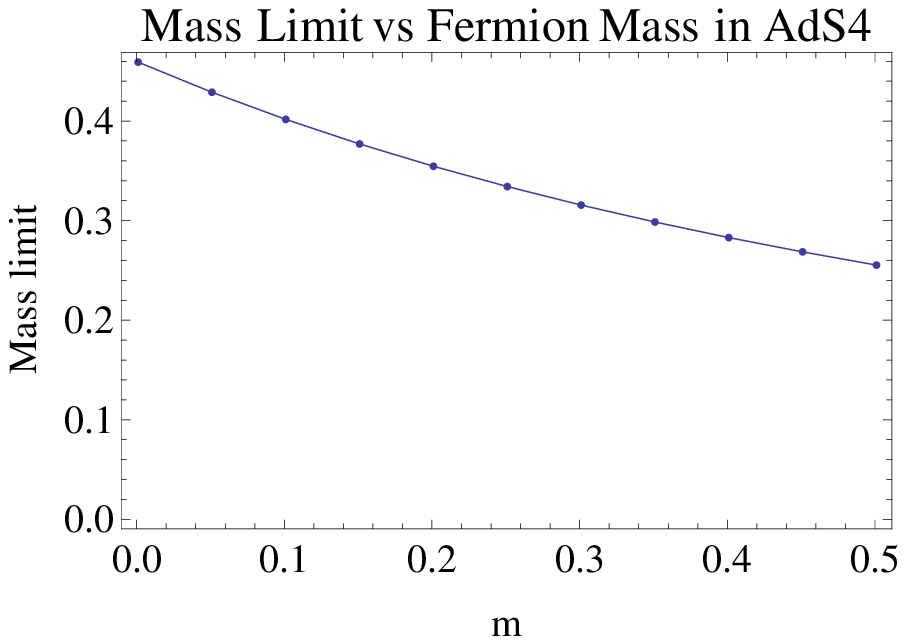}
        \includegraphics[scale=0.9]{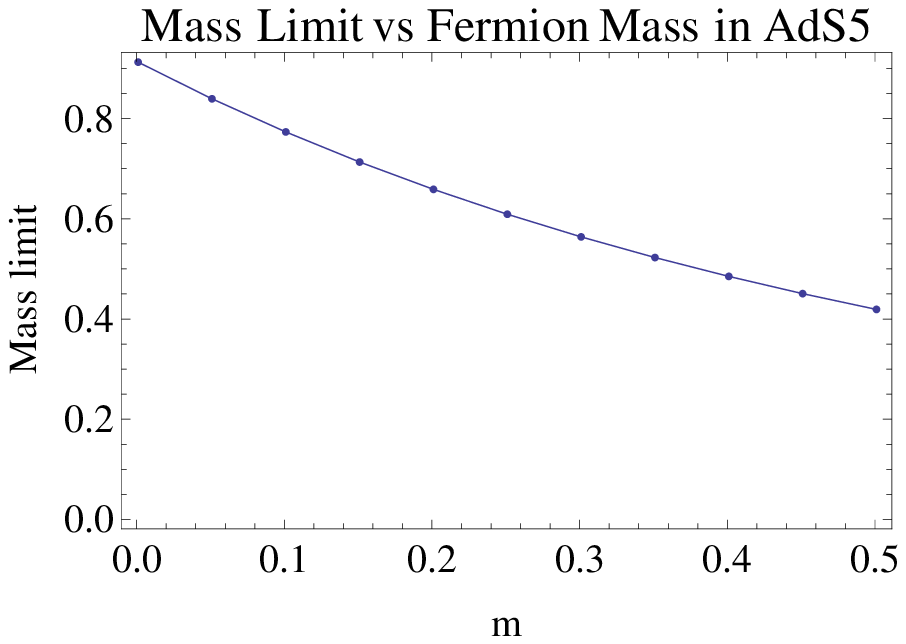}
        \caption{Relation between mass limit and fermion mass in $AdS_{4}$ and $AdS_{5}$.} \label{Mvsm}
    \end{figure}

\begin{figure}
        \includegraphics[scale=0.9]{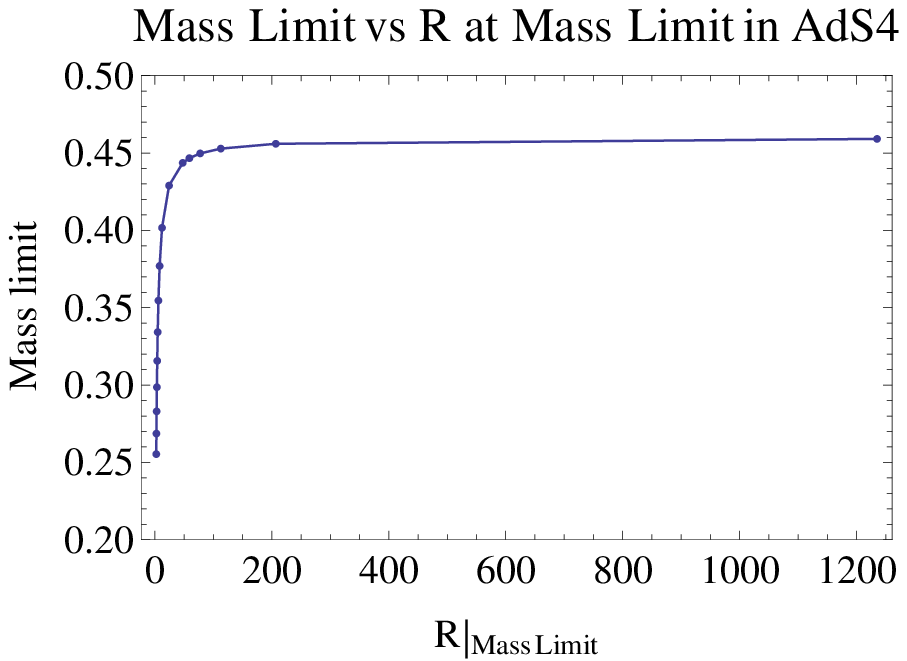}
        \includegraphics[scale=0.9]{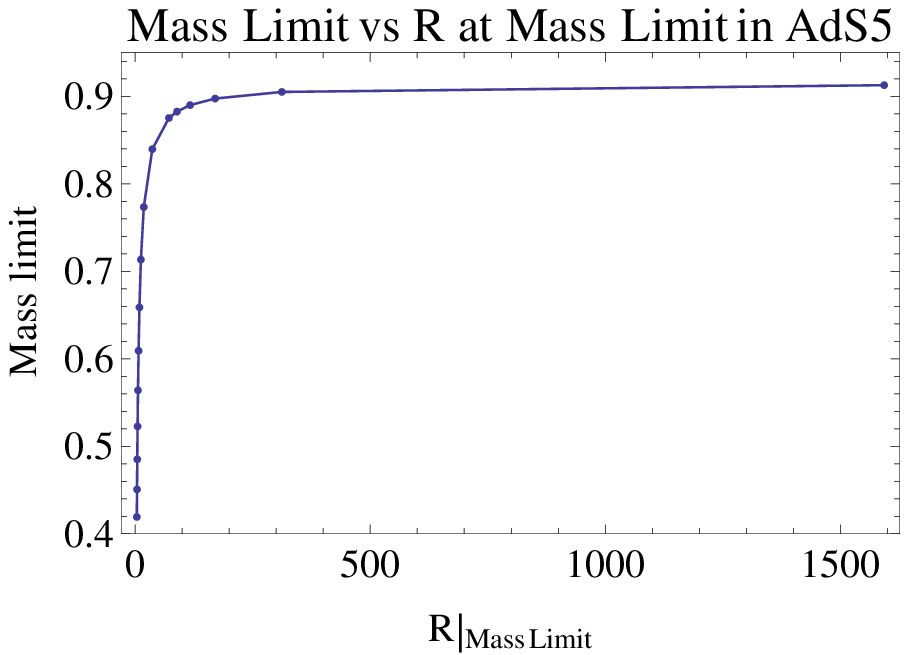}
        \caption{Relation between mass limit and radius in $AdS_{4}$ and $AdS_{5}$.}\label{MvsR}
    \end{figure}

In this section, we will discuss the relationship between the mass limit, radius, and mass of the fermions in the AdS star as a result of the numerical solutions to the TOV equations.  We have chosen $g_{f}=2$ for all of the plots except stated otherwise~(plots involving binding energy have $g_{f}=1$).  First, the mass limit of the AdS star is found to be a decreasing function of $m$ as shown in Fig.~\ref{Mvsm}.  A quantum property of fermion gas which determines the mass limit of the star is the degenerate pressure.  Even massive fermions at zero temperature are forced to possess kinetic energy due to exclusion principle and thus they inevitably exert pressure against the gravity.  As $m\to 0$, most of the energy of the fermions becomes kinetic, resulting in the highest possible degenerate pressure and mass limit.  Higher pressure also implies bigger star as we can see from Fig.~\ref{MvsR}.  Another crucial feature is that the mass limit is not sensitive to $m$ when $m$ is small as we can see from the saturation of the mass limit for $m\to 0$ and ever increasing $R$.

Figure~\ref{MvsR} shows the relation between mass limit and radius of AdS stars in four and five dimensions.  By varying the fermion mass $m=0.001-0.5$, we numerically obtain the mass limit and the corresponding radius of the AdS star at each $m$.  AdS star with large $m$ has lower mass limit and smaller radius.  There is a possibility for $M<m$ as $m$ increases.  It is originated from the choice of units when we set $G=1, \ell=1$.  For a fixed $m, G\ell^{2}$; $\ell$ can always be chosen to be so large, $G$ to be so small that $M~\ell^{d-1}$ becomes larger than $m$ and vice versa.  As $m\to 0$, the radius continues to grow but the mass limit saturates at an upper bound mass.  For $AdS_{4,5}$, the upper bound $M_{\rm limit}(m\to 0)=0.459,0.913$ respectively. 

Accordingly on the boundary, the conformal dimension of the corresponding composite operator has an upper bound at $M_{\rm limit}\ell$ in the relativistic fermion limit.  The lesser the conformal dimension of the constituent fermionic operator $\Psi$, the higher the conformal dimension of the composite $\Phi$ before the deconfinement occurs.  Existence of the upper bound implies that the deconfinement is inevitable provided that the energy density~(and consequently $\Delta$) is sufficiently large.  Injecting mass to the bulk until it exceeds the mass limit and the AdS star inevitably collapses corresponds to the deconfinement of the CFT matter on the AdS boundary when the energy density exceeds a critical value~(larger and larger number of $\Psi$ accumulates to form $\Phi$ until reaching critical value).

\subsection{Relations between Fermi energy and fermion mass}

\begin{figure}
        \includegraphics[scale=0.9]{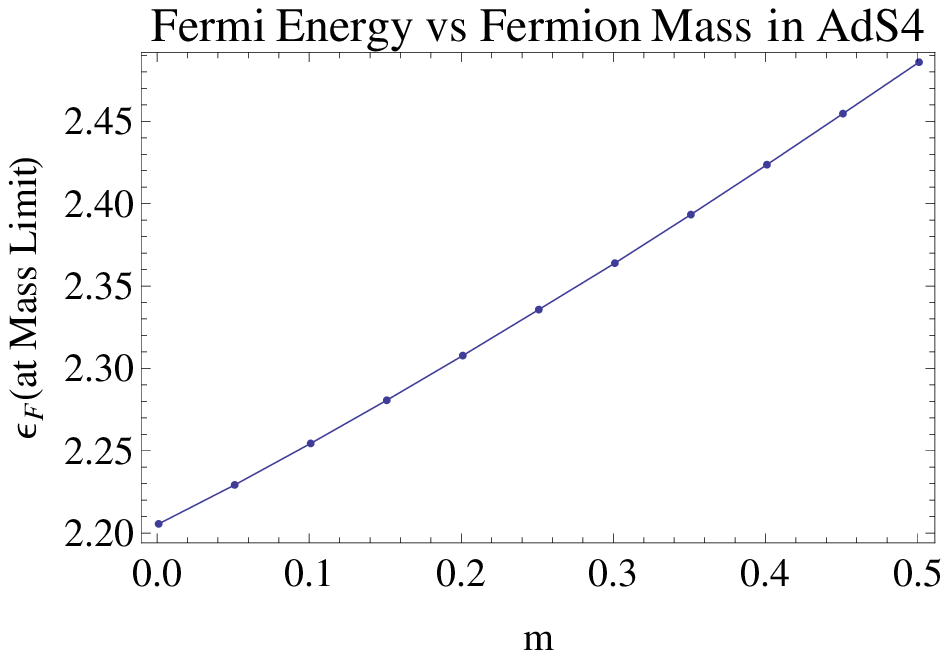}
        \includegraphics[scale=0.9]{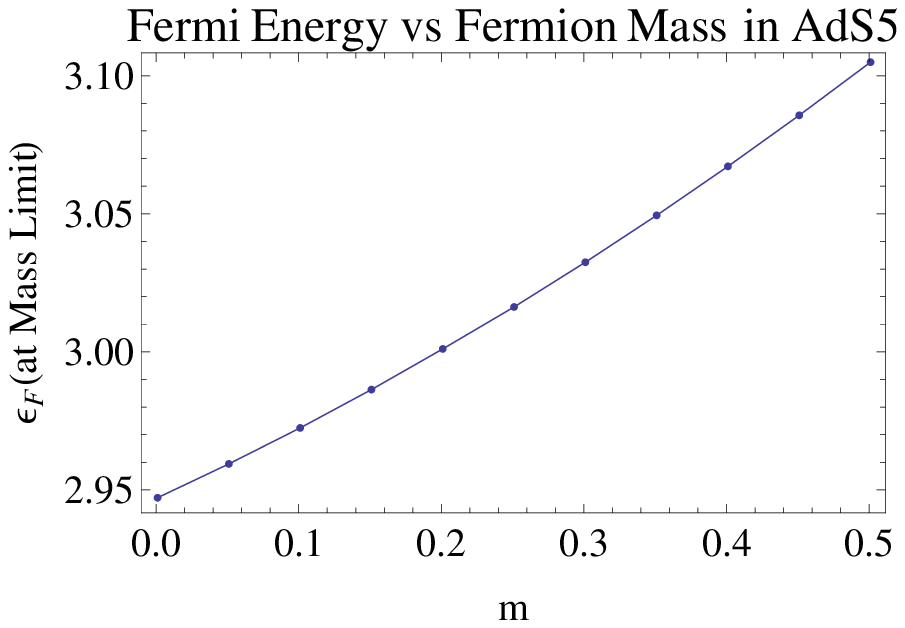}
        \caption{Relation between Fermi energy and fermion mass in $AdS_{4}$ and $AdS_{5}$.} \label{efvsm}
    \end{figure}

At center of the star, gravitational force from both the AdS background and the self-gravity become zero.  Therefore the chemical potential at $r=0$ is simply the Fermi energy $\epsilon_{F}$.  The chemical potential gets redshifted as $r$ increases and becomes $m$ at the surface of the star.  Each class of star with a given value of $m$ has a mass limit at a particular $\mu(r=0)=\epsilon_{F}$.  Numerical results are presented in Fig.~\ref{efvsm}.  The Fermi energy or central chemical potential at the mass limit is found to be an increasing function of the fermion mass.  

On the boundary, this result is consistent with relation $\epsilon_{F}\ell=(\Delta_{0}+n_{F})$.  The dependence of $\epsilon_{F}$ and $m$ is almost linear at the critical $\Delta=M_{\rm limit}\ell$ where deconfinement occurs.  We can interpret that $n_{F}$ at the deconfinement is also a slowly increasing function of $\Delta_{0}\simeq m\ell$.

\subsection{Binding energy}

\begin{figure}
        \includegraphics[scale=0.9]{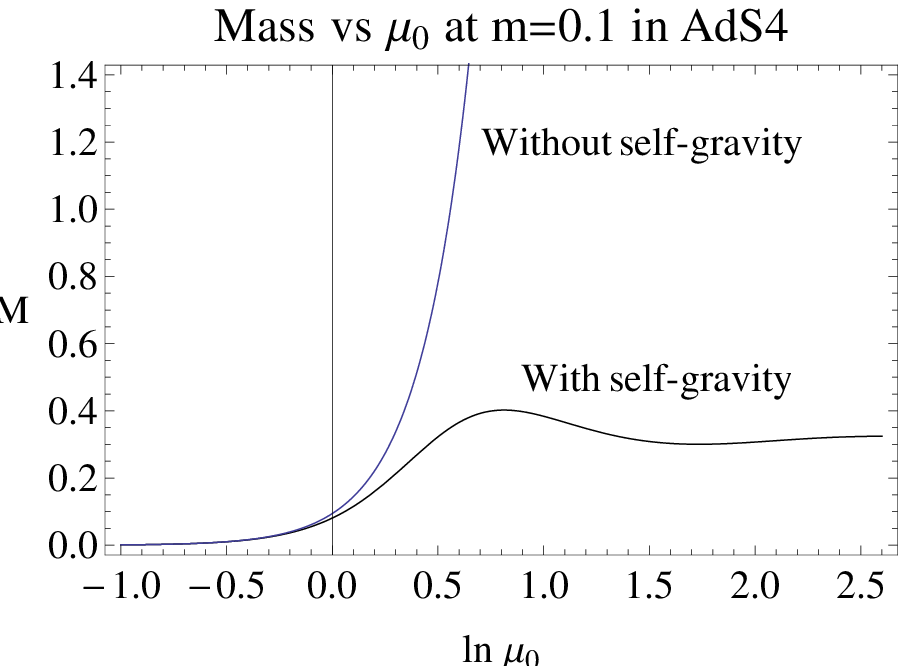}
        \includegraphics[scale=0.9]{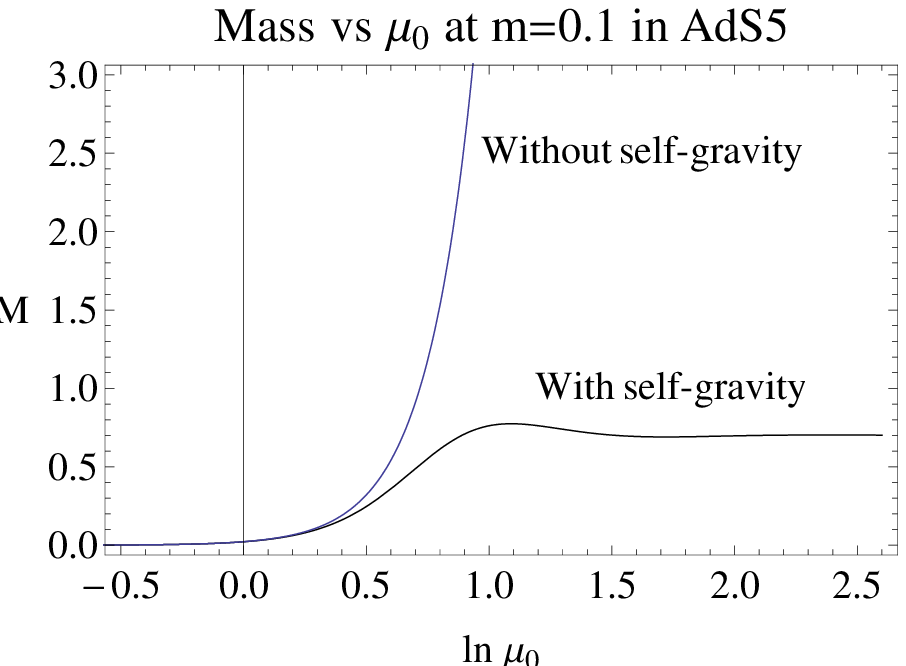}
        \caption{Mass limit curve of AdS stars for fermion mass $m = 0.1$ in $AdS_{4}$ and $AdS_{5}$.  Note that $g_{f}=1$ for these plots.} \label{befig}
    \end{figure}

\begin{figure}[h]
 \centering
        \subfigure[]{\includegraphics[width=0.45\textwidth]{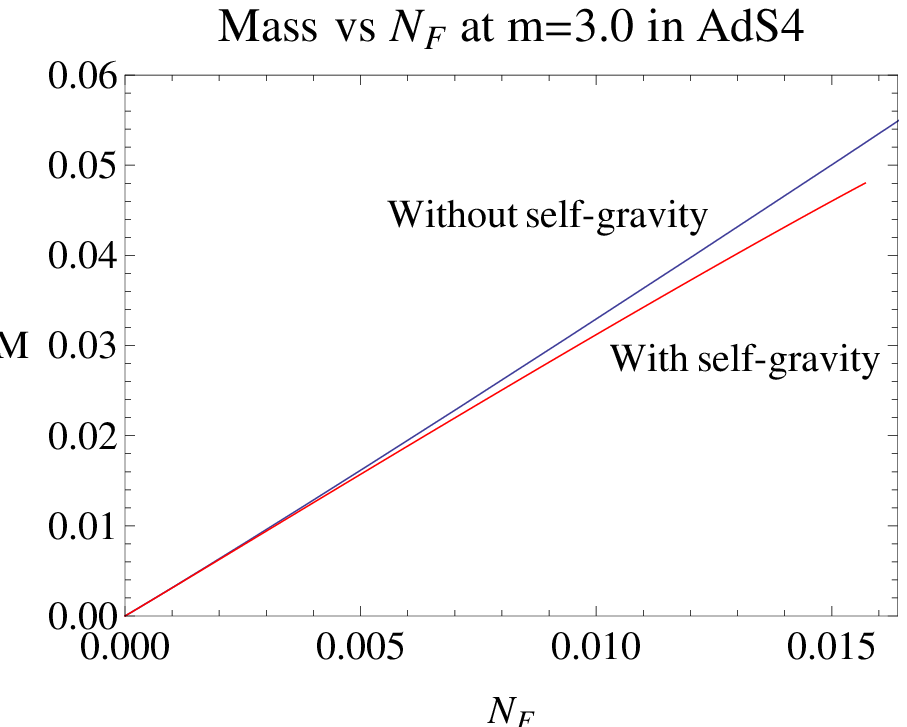}}\hfill
        \subfigure[]{\includegraphics[width=0.48\textwidth]{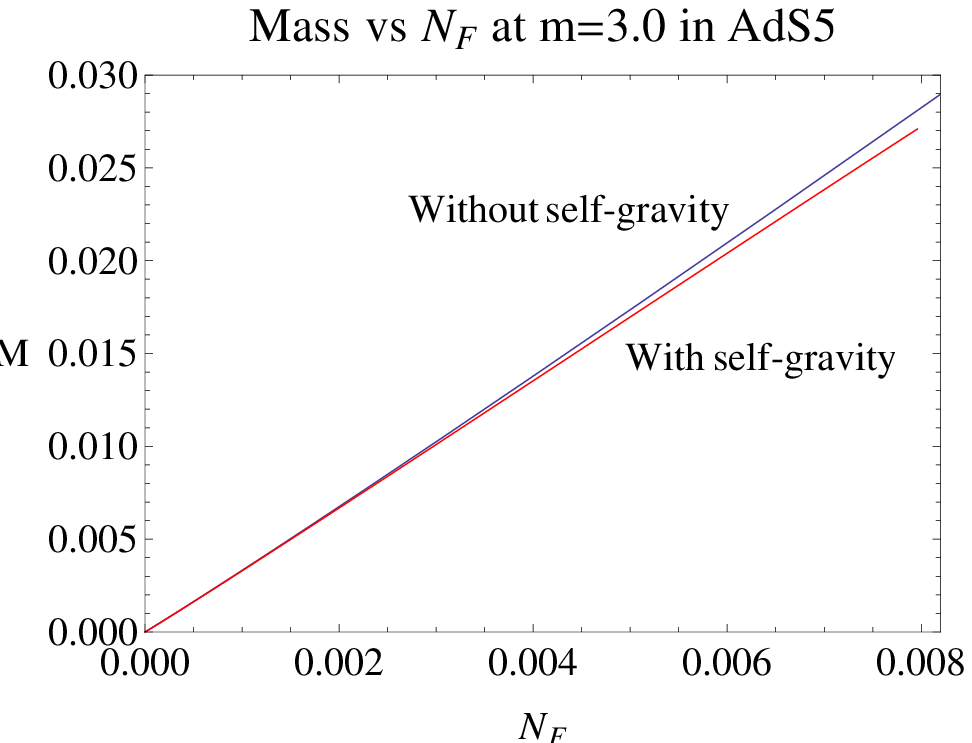}}\\
        \subfigure[]{\includegraphics[width=0.5\textwidth]{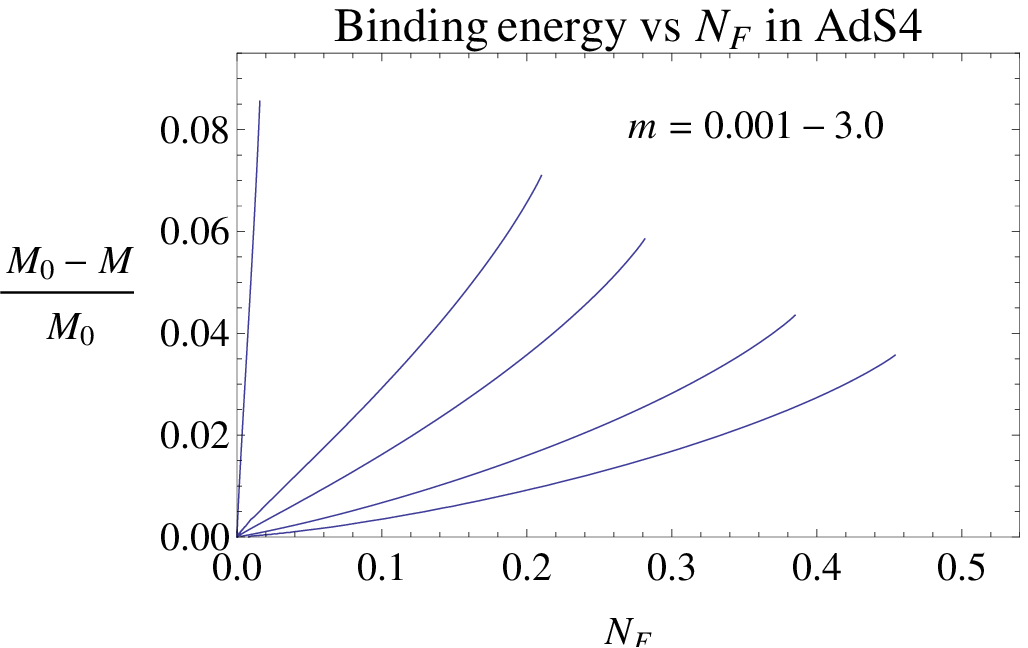}}\hfill
        \subfigure[]{\includegraphics[width=0.5\textwidth]{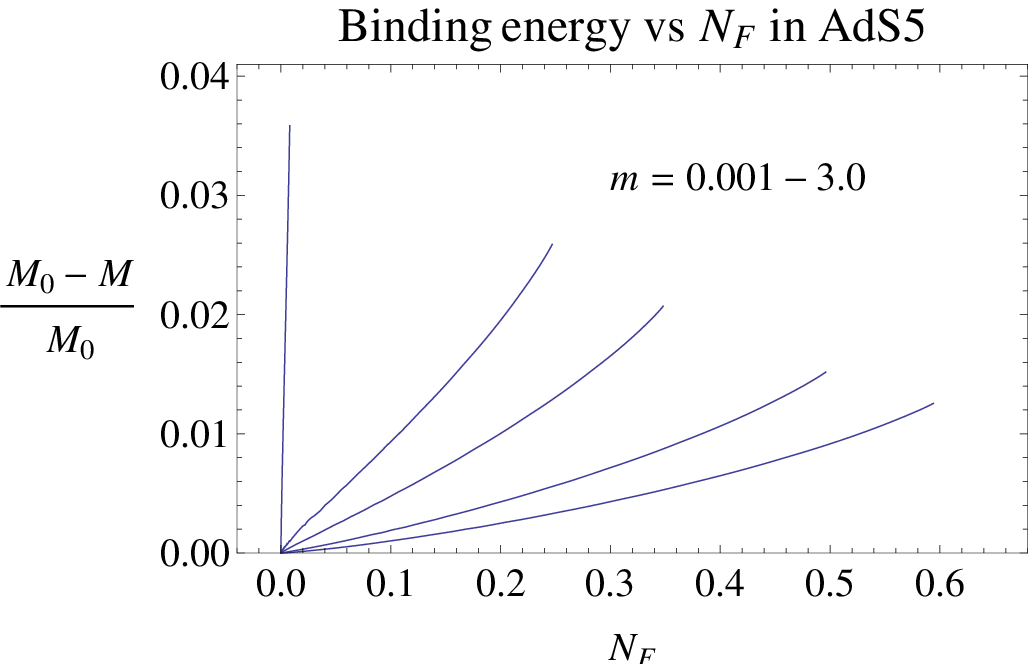}}        
        \caption{(a) and (b): Mass and total particle number of the AdS stars for $m=3.0$.  The curves with self-gravity are truncated at the mass limit.  (c) and (d): Binding energy of AdS stars for fermion mass $m = 0.001, 0.1, 0.3, 0.5, 3.0$~(lower to upper curves respectively) in $AdS_{4}$ and $AdS_{5}$.  Note that $g_{f}=1$ for all plots. } \label{befig1}
\end{figure}

The binding energy of the AdS star can be calculated from the difference between the mass with and without the self-gravity at the fixed total particle number.  We can numerically calculate the mass of the AdS star from the coupled equations (\ref{EoM of M}), (\ref{EoM of mu}) in the presence of self-gravity, and analytically from Eqn.~(\ref{mat0}) when self-gravity is neglected.  However, these masses are determined at a fixed $\mu_{0}$~(Fig.~\ref{befig}), consequently they correspond to different particle numbers.  In order to obtain the binding energy of the star, we need to subtract the masses at a fixed particle number $N_{F}$~(Fig.~\ref{befig1} (a) and (b)).

From Fig.~\ref{befig1} (c) and (d), the binding energy fraction is smaller for smaller fermion mass $m$.  As the fermion mass approaches zero, the binding energy saturates.  We can also see the saturation phenomenon in Fig.~\ref{MvsR}~(as the radius grows with diminishing $m$) and Fig.~\ref{Mvsm}.  The implication for the conformal physics on the boundary is the following.  For operator $\Psi$ with small conformal dimension $\Delta_{0}$, the deconfinement transition occurs when the conformal dimension of the composite operator $\Phi$ is relatively larger than when $\Delta_{0}$ is large.  However, even when $\Delta_{0}$ becomes very small~($m\to 0$ in the bulk) the deconfinement still occurs at a saturated value of $\Delta$, the conformal dimension of the composite $\Phi$.  There is an upper bound on the critical conformal dimension $\Delta_{c}$~(corresponding to $M_{\rm limit}$) above which deconfinement is inevitable.  For a given fermion mass $m$, the mass limit of the AdS star would correspond to the critical conformal dimension $\Delta_{c}\equiv \Delta_{c,0} + \Delta_{int.}$ of the operator $\Phi$ while $\Delta_{c,0}$ would be given by the mass of the AdS star neglecting self-gravity at the same total particle number, $M_{0}({\rm same}N_{F})$.  Namely, we have the following relations
\begin{eqnarray}
M_{0} \ell & = & \Delta, \\
M_{\rm limit}\ell & =& \Delta + \Delta_{int.}.  \label{Dlim}
\end{eqnarray}
From Fig.~\ref{MvsR},{\ref{Mvsm}} and {\ref{befig1}}~($\ell = 1$), the estimated values of the critical conformal dimensions are 
\begin{eqnarray}
\Delta_{c}(m\to 0) & \simeq & 0.459, 0.913  \quad ({\rm for} AdS_{4,5}),  \\
\Delta_{c,0} & \simeq & \frac{\Delta_{c}}{1-\alpha},  \\
& = &  0.476, 0.924,  \quad ({\rm for} AdS_{4,5}),
\end{eqnarray}
where we have used the binding energy fraction at the mass limit $\alpha \equiv (M_{0}-M)/M_{0}(m\to 0)=0.0358, 0.0125$ for $d=4, 5$ respectively.  The quantity $\Delta_{c}(m\to 0)$ is the upper bound for $\Delta_{c}$ at other arbitrary fermion masses.  Certainly in the zero fermion mass limit, the corresponding conformal dimension is given by Eqn.~(\ref{mdcorr}), $\Delta_{0}(m\to 0)=d$, not zero while the latter two relations of Eqn.~(\ref{corr}) remain valid.    

\subsection{Relation between mass limit and total particle number}

\begin{figure}
        \includegraphics[scale=0.9]{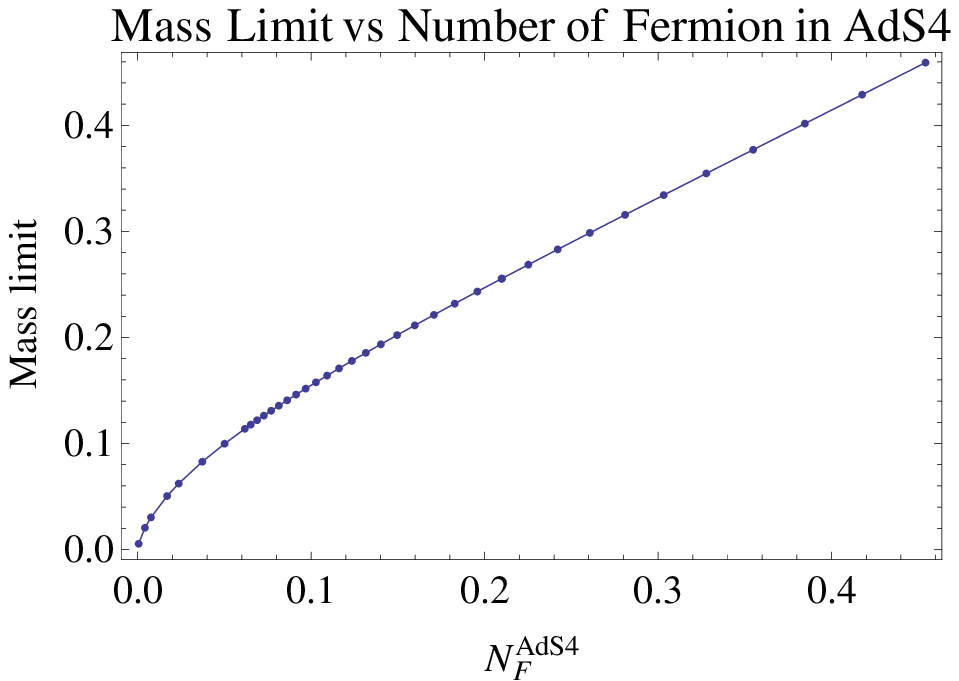}
        \includegraphics[scale=0.9]{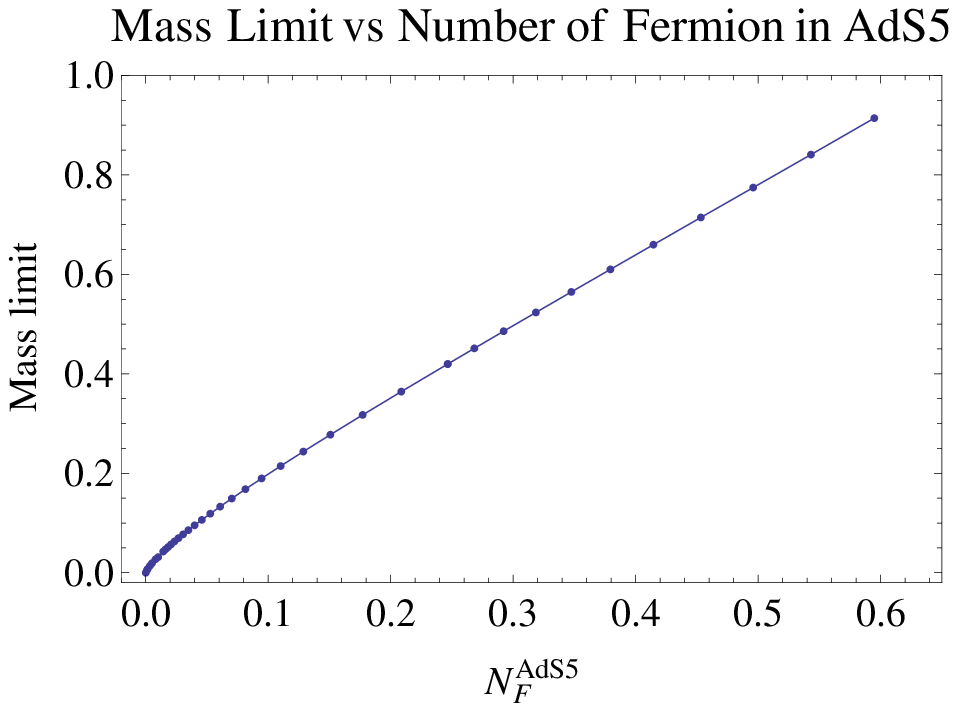}
        \caption{Mass limit and total particle number at the mass limit for varying $m$.  As $m\to 0$, the relationship converges to a linear function.  The mass and particle number reach maximal values at $m=0$.   } \label{mlimn}
    \end{figure}

An interesting result is found between the mass limit and corresponding total particle number as is shown in Fig.~\ref{mlimn}.  For sufficiently small $m$, the relation between $M_{\rm limit}$ and $N_{F}$ converges to a linear function for both $AdS_{4,5}$.  For larger $m$, both $M_{\rm limit}$ and $N_{F}$ approach zero.  This result is intriguing considering the fact that for small $m$, most fermions are kinetical and the rest energy is negligible.  We do not expect the total mass to be proportional to the number of particle especially when the chemical potential is redshifted along the radial coordinate, see e.g. Fig.~\ref{befig1}~(a), (b)~(see also Fig.~8 of Ref.~\cite{Arsiwalla:2010bt}).  Numerically, we can understand this result from the behaviour of binding energy shown in Fig.~\ref{befig1}.  As $m\to 0$, the binding energy fraction becomes negligible, therefore the expression of mass and particle number without self-gravity given by Eqn.~(\ref{mat0}), (\ref{ntot}) can be used.  Setting $m=0$, we obtain the linear relation $M=(d-1)N_{F}/d$.  

On the boundary, we can understand this behaviour in the large $\ell$ limit as discussed in Section \ref{lell}.  Since $\Delta \sim c\sim \ell^{d}, \Delta_{0}\sim \ell, N_{F}\sim \ell^{d-1}$, hence both $M_{\rm limit}$ and $N_{F}$ are of order $\ell^{d-1}$.  Additionally, $\Delta, N_{F}$ are not sensitive to small $m$ in the double scaling limit as is also confirmed by the bulk result in Fig.~\ref{MvsR}.  Consequently, we expect the relation between $M_{\rm limit}$ and $N_{F}$ to be linear when $\ell$ is large and $m$ is small.  This result remains valid for smaller $\ell$ and larger $G$ as long as $G\ell^{2}$ is fixed as we can see from Fig.~\ref{mlimn}.  

\begin{figure}
        \includegraphics[scale=0.9]{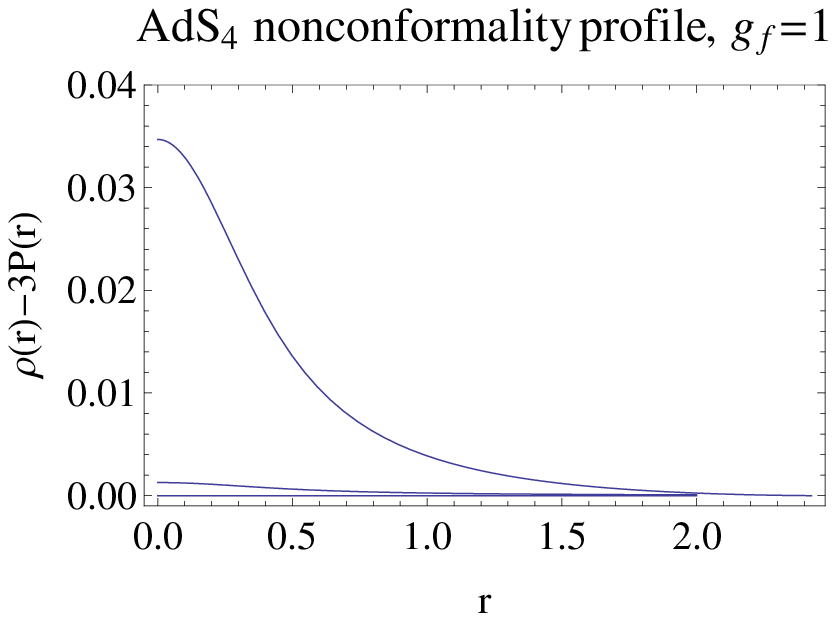}
        \includegraphics[scale=0.9]{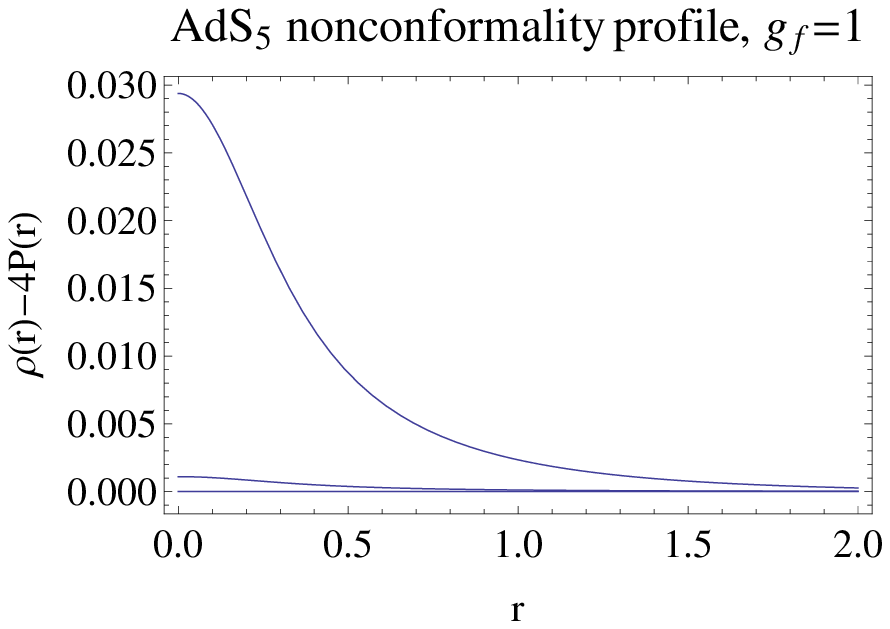}
        \caption{Conformal breaking factor profile $\rho - (d-1) P$ for $m=0.001, 0.1, 0.5$~(lower to upper curve) of the AdS stars at the mass limit in $d= 4, 5$.} \label{confact}
    \end{figure}

Lastly, let us comment on the ${\it bulk}$ conformal limit, $m\to 0$ of the AdS star.  As expected from the linear equation of state, Eqn.~(\ref{zerom}), the conformal breaking factor $\rho - (d-1) P$ approaches zero as $m\to 0$.  This is shown in Figure~\ref{confact}.  
In this bulk conformal limit, the relation between the mass limit and the total particle number~(at the mass limit) becomes linear as is shown in Fig.~\ref{mlimn}.  The relationship is not obvious due to the extra $B(r)$ in the definition of $N_{F}$ comparing to $M$.

\subsection{$g_{f}$ factor and internal degrees of freedom}

    \begin{figure}
        \includegraphics[scale=0.9]{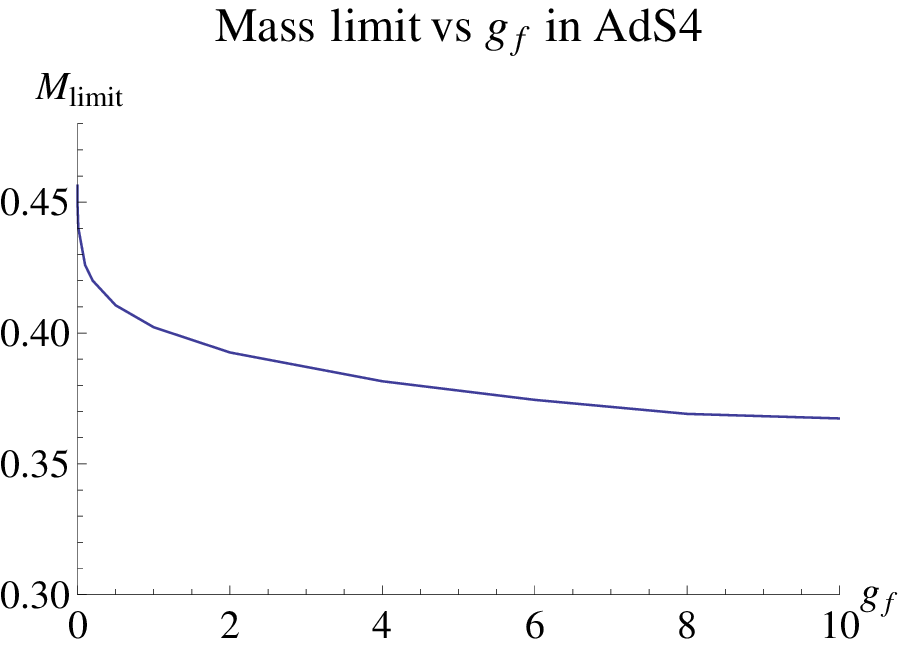}
        \includegraphics[scale=0.9]{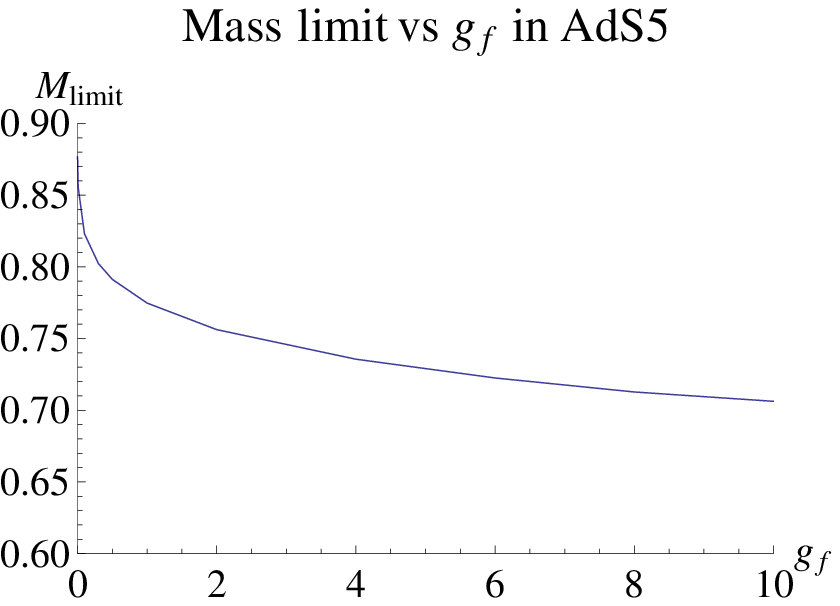}
        \caption{Relation between mass limit and spin factor representing the number of internal degrees of freedom of the fermion for $m=0.1$.} \label{mgs}
    \end{figure}

The spin degrees of freedom can be included in the analysis as a $g_{f}$ factor multiplying to the density and pressure of the fermion content of the star.  Incidentally, this factor can actually represent all other internal degrees of freedom of the bulk fermion such as the colour and flavour as long as a universal mass~(and chemical potential) is assumed among those degrees of freedom.  The dependence on the number of internal degrees of freedom is shown in Fig.~\ref{mgs}.  The mass limit is found to be a slowly decreasing function of $g_{f}$ even though the pressure and density are increased.  Such paradoxical behaviour can be understood by considering the equation of motion of the chemical potential, Eqn.~(\ref{EoM of mu}).  Increasing both the density and pressure with a factor $g_{f}$ will make $\mu(r)$ decreases more rapidly with respect to $r$, consequently the star radius becomes smaller making the accumulated mass smaller.  We also found that the central chemical potential $\mu_{0}$ at the mass limit decreases with increasing $g_{f}$ for a fixed $m$. 

For both bulk and boundary picture, $g_{f}$ represents the same localized degrees of freedom.  If there are more colour and flavour degeneracy of CFT ``nucleon"~(or more generically ``multiquark"), the deconfinement could occur at slightly lower $\Delta$~(and lower energy density) according to Fig.~\ref{mgs}.

\section{Small and large AdS black holes}   \label{IV}

In this section, we briefly review the thermodynamics of black holes in the AdS space.  We will present the results in general dimension.  The analysis is based on the Euclideanized time action which neglects the effects of self-gravity of the radiation in the AdS space.    

There are two possibilities of static black holes in the AdS space, a small and large one~\cite{Hawking:1982dh}.  A small black hole~(SBH) is defined to be the branch which the Hawking temperature decreases as the black hole mass grows and thus it has a negative heat capacity.  A large AdS black hole~(LBH), on the other hand, is the black hole in the positive heat capacity branch with the Hawking temperature an increasing function of the black hole mass.  Quantum fluctuations near the horizon of a black hole generate Hawking radiation at a fixed temperature which is related to the Bekenstein entropy in a standard thermodynamical manner.  The heat capacity of a black hole can be defined once its temperature is determined.  

Starting with the classical action of AdS-Schwarzschild spacetime in $d$ dimension,
\begin{eqnarray}
I & = & -\frac{1}{16\pi G}\int~d^{n+1}x\sqrt{-g}\left( R + \frac{n(n-1)}{\ell^{2}}\right), \label{gact} \\
& = & \frac{n}{8\pi G\ell^{2}}\int~d^{n+1}x\sqrt{-g},
\end{eqnarray}
proportional to the volume of the spacetime.  In the presence of a black hole, we set $\chi(r)=0, M(r)=M$ in Eqn.~(\ref{bmetric}) for the metric.  Both the volume factor of the AdS and the AdS-Schwarzschild spacetime are infinite but the difference is finite, the regulated action obtained by substracting the two at the same asymptotic radius is~\cite{Witten:1998zw} 
\begin{eqnarray}
I_{reg}& = & \frac{V_{n-1}}{4G}~\frac{(\ell^{2}-r_{+}^{2})~r_{+}^{n-1}}{n ~r_{+}^{2}+(n-2)\ell^{2}}, \label{Icl}
\end{eqnarray}
where $n=d-1$ is the dimension of the boundary spacetime and $r_{+}$ is the horizon radius of the black hole.  The cosmological constant is related to the AdS radius $\ell$ by
\begin{eqnarray}
\Lambda & = & -\frac{n(n-1)}{2~\ell^{2}}.
\end{eqnarray}
The corresponding Hawking temperature $T_{H}=1/\beta$ is a function of the horizon radius as
\begin{eqnarray}
\beta & = & \frac{4\pi\ell^{2}r_{+}}{n ~r_{+}^{2}+(n-2)\ell^{2}}.  \label{hor}
\end{eqnarray}
The total energy~(mass) of the AdS-BH can be calculated thermodynamically to be
\begin{eqnarray}
E & = & \frac{\partial I}{\partial \beta} = \frac{V_{n-1}(n-1)}{4G}\frac{(\ell^{2}+r_{+}^{2})~r_{+}^{n-2}}{4\pi~\ell^{2}},  \label{BHM}
\end{eqnarray}
with the heat capacity
\begin{eqnarray}
C_{V}& = & \frac{V_{n-1}}{4 G}\frac{(n-1)r_{+}^{n-1}(n ~r_{+}^{2}+(n-2)\ell^{2})}{n ~r_{+}^{2} - (n-2)\ell^{2}}.  \label{cvbh}
\end{eqnarray}
The heat capacity is negative when $r_{+}<\ell \sqrt{(n-2)/n}\equiv r_{c}$ and positive for $r_{+}>\ell \sqrt{(n-2)/n}$ respectively.  As a demonstration of the holographic principle, we also calculate the entropy of the AdS-BH,
\begin{eqnarray}
S &=& \beta E - I = \frac{V_{n-1}}{4 G}r_{+}^{n-1} = \frac{A_{n-1}(r_{+})}{4G},
\end{eqnarray}
where $A_{n-1}(r_{+})$ is area of the event horizon of the black hole.  

\begin{figure}
      \centering  
	\includegraphics[]{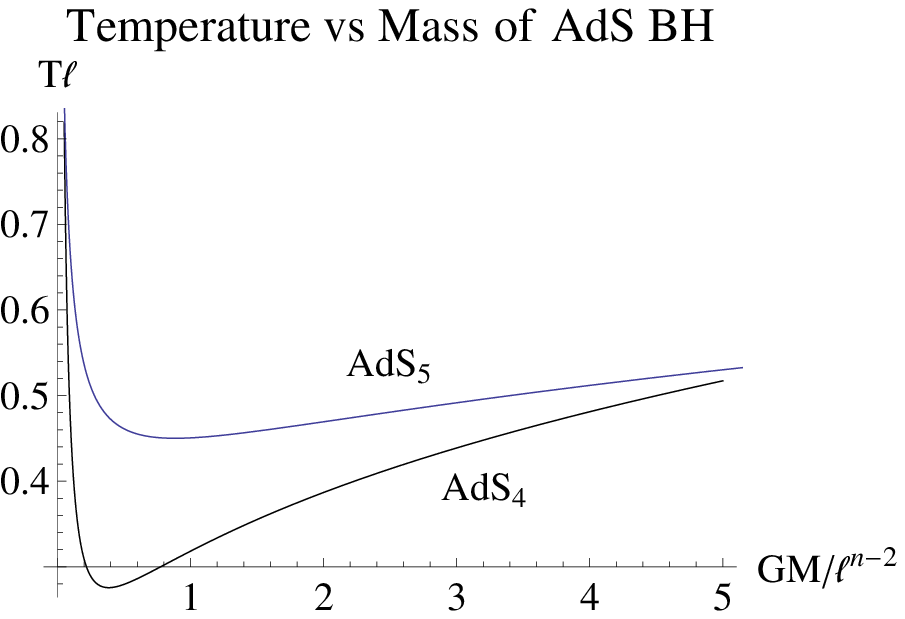}
        \caption{Temperature versus mass of the AdS-BH.}  \label{bhbr}
    \end{figure}

The minimal Hawking temperature occurs when $r_{+}=r_{c},$
\begin{equation}
 T_{\rm min}=\frac{\sqrt{n(n-2)}}{2\pi\ell} = \frac{1}{2\pi}\sqrt{-\frac{2(n-2)\Lambda}{n-1}}. 
\label{tmin} 
\end{equation}
For $d=4,5~(n=3,4)$, this is equal to $\sqrt{-\Lambda}/2\pi, \sqrt{-4\Lambda/3}/2\pi$ respectively.  Above the minimal temperature, AdS-BH branches into small and large ones with negative and positive heat capacity as shown in Fig.~\ref{bhbr}.  
The critical mass distinguishing small and large black holes is at $r_{+}=r_{c}$
\begin{eqnarray}
M_{c}& = & \frac{V_{n-1}\ell^{n-2}}{8\pi G}\frac{(n-1)^{2}}{n} \left( \frac{n-2}{n}\right)^{(n-2)/2}  \label{bhmc}
\end{eqnarray}
In $d = 4, 5~(n=3, 4)$, the critical mass is \[ \frac{2 ~\ell}{3\sqrt{3} ~G}, \frac{9\pi ~\ell^{2}}{32 ~G}
\]
or $0.385, 0.884$ for $\ell, G=1$ respectively.  Classically, both small and large black holes can exist in the AdS space in any dimensions.  Semiclassically, however, thermal fields in small~(large) AdS-BH spacetime is thermodynamically less~(more) preferred than pure radiation in the AdS space.  The free energy of the fields in the AdS-Schwarzschild background at finite temperature could be identified with the gravity action in the Euclideanized time.  From Eqn.~(\ref{Icl}), (\ref{hor}), the regulated gravity action is related to the temperature $T=1/\beta$ by
\begin{eqnarray}
I_{\pm} & = & \frac{V_{n-1}}{4G}\frac{\beta}{4\pi}\left( \frac{4\pi \ell^{2}}{2n}\right)^{n-2}\Bigg( 1- \beta_{c}^{2}\frac{n-2}{n}\left(\frac{1}{\beta}\pm\sqrt{\frac{1}{\beta^{2}}-\frac{1}{\beta_{c}^{2}}}\right)^{2}\Bigg)\left(\frac{1}{\beta}\pm\sqrt{\frac{1}{\beta^{2}}-\frac{1}{\beta_{c}^{2}}}\right)^{n-2}  \label{IBH}
\end{eqnarray}
where $\beta_{c}\equiv 2\pi \ell/\sqrt{n(n-2)}$ and the $+(-)$ sign corresponds to the LBH(SBH) respectively.  The free energy of the AdS-BH can be calculated from the partition function $F=-T\log Z= TI$,
\begin{eqnarray}
F &=& \frac{V_{n-1}r_{+}^{n-2}}{16\pi G}\left( 1-\frac{r_{+}^{2}}{\ell^{2}}\right), \label{freeE}
\end{eqnarray}
which is positive~(negative) when $r_{+}<(>) \ell$.  At $r_{+}=\ell$, the BH temperature is $(n-1)/2\pi\ell\equiv T_{1}>T_{\rm min}$.  By comparing the free energy of the BH in thermal equilibrium with radiation and the thermal AdS without self-gravity, we thus conclude that for 
\begin{equation}
T_{\rm min}<T<T_{1},  
\end{equation}
the BH~(regardless of the size) in thermal equilibrium is less preferred thermodynamically than the pure thermal AdS without self-gravity.  For higher temperature $T>T_{1}$~(the Hawking-Page temperature), the free energy of the LBH~(SBH) becomes negative~(positive).  LBH becomes more preferred than the pure thermal radiation while the SBH is the least preferred.  Note that the self-gravity of the radiation has been neglected in this consideration for both thermal AdS and BH in thermal equilibrium.

\section{Self-gravity effects of the radiation and the bulk conformal limit}  \label{V}

In this section we consider effects of self-gravity of the radiation and matter in the AdS space.  For AdS space without BH, this is simply the ``radiation" star with infinite radius.  For the star of massless fermion, the AdS space behaves like a box.  As a consequence, the fermions develop nonzero chemical potential even at zero temperature within the star.  As we have numerically demonstrated, the fermionic AdS star in $m\to 0$ limit at zero temperature has a finite mass limit with an infinite radius!  In this sense, it is a ``radiation" star at zero temperature.  On the other hand, the conventional radiation star with zero chemical potential~(and thus nonzero temperature) in the AdS space has been studied a couple of times, e.g. Ref.~\cite{Page:1985em},\cite{Vaganov:2007at}.  Starting from the number density of the massless particle specie $i$ given by
\begin{eqnarray}
n_{i}& = & \int\frac{d^{n}\vec{p_{i}}}{(2\pi)^{n}}(e^{\beta E_{i}}\pm 1)^{-1}
\end{eqnarray}
where the sign is $+(-)$ for fermion(boson), we obtain
\begin{eqnarray}
\rho^{b}& = & \frac{T^{n+1}}{2^{n-1}\pi^{n/2}}\frac{\Gamma(1+n)}{\Gamma(n/2)}\zeta(1+n) \label{radrp}
\end{eqnarray}
for boson and $\rho^{f}=(1-2^{-n})\rho^{b}$ for fermion.  The total energy density of radiation is thus
\begin{eqnarray}
\rho & = &  g^{b}\rho^{b}+g^{f}\rho^{f}\equiv g_{\rm eff}\rho^{b}\equiv g_{*}T^{n+1}, \label{radrp1}
\end{eqnarray}
summing over all species of particle and the radiation pressure is given by $P=\rho/(d-1)$.  It has been shown in Ref.~\cite{Page:1985em},\cite{Vaganov:2007at} that the radiation star in the AdS space even with infinite radius generically has mass limit above which gravitational collapse to BH is inevitable.  The mass limit corresponds to the critical maximal temperature of the radiation at the center, $r=0$ of the AdS space.  The radiation with central temperature higher than this value is unstable under radial perturbation.  

It is interesting to compare the mass limit obtained from the conventional radiation star and the mass limit with $m\to 0$ obtained in Section \ref{adsself}.  The former has $\mu = 0, T>0$ while the latter has $T=0, \mu > 0$.  As is demonstrated in Appendix A of Ref.~\cite{Burikham:2012kn}~(Eqn.~(A.32)), the equation of motion of $T(r)$ and $\mu(r)$ are identical, i.e. $T'/T=\mu'/\mu = -A'/A$.  Therefore, there is a symmetry under $T$ and $\mu$ exchange in the system if the density $\rho$ and pressure $P$ have the same dependence between $T$ and $\mu$.  This is exactly the case when $m=0$, the bulk conformal symmetry demands that $\rho, P$ can only depend on a single dimensionful quantity, either $T$ or $\mu$.  Dimensional analysis implies that $\rho, P$ will have exactly the same independence on $T$ or $\mu$; for conventional radiation star, they are given by Eqn.~(\ref{radrp}), (\ref{radrp1}), for fermionic star with $m=0$, they are given by Eqn.~(\ref{zerom}).  Consequently, {\it both cases will give exactly the same mass limit}!  This remarkable result implies that the upper mass limit we found in Section \ref{adsself} is actually the same as the mass limit of the radiation in the AdS space with the self-gravity, a genuine {\it universal} mass limit of the AdS space.  

Numerically, we found that the critical central temperature at the mass limit~($g_{*}=2$) of the thermal AdS space is $0.622(0.776)$ for $AdS_{4(5)}$ respectively.  The mass limit is $0.459(0.913)$ for $AdS_{4(5)}$, exactly the same as the mass limit of the fermionic star in Section \ref{adsself} as expected from the symmetry between $T$ and $\mu$ of the two cases in the bulk conformal limit.  Another remarkable behaviour is the scaling invariance of the mass limit with respect to $T$ or $\mu$.  Since the expression of $\rho$ and $P$ does not change form under rescaling $T$ or $\mu$~(due to bulk conformal symmetry), the mass limit curve only shifts in the $T$ or $\mu$ direction under the rescaling.  The mass limit remains the same.  As a consequence, the mass limit does NOT depend on $g_{*}$~(or $g_{f}$) since we can rescale $T\to T g_{*}^{1/d}$~(or $\mu$) to eliminate $g_{*}$ from the equation of motion.  Certainly, the critical $T$ or $\mu$ at the mass limit will be shifted with a factor $g_{*}^{-1/d}$ accordingly.

On the boundary, the universal mass limit can be interpreted to be the dual of universal critical temperature of the strongly coupled CFT matter above which deconfinement should occur.  The critical temperature in the gauge picture is the Hawking temperature of the AdS-BH with mass equal to the universal mass limit of the AdS star given by Eqn.~(\ref{hor}), (\ref{BHM})~(see also Fig.~\ref{bhbr}).  For $AdS_{4,5}$,
\begin{eqnarray}  
T_{crit}& = & 0.277, 0.450   \label{Tcrit}
\end{eqnarray}
respectively.  Note that $T_{crit}$ is slightly larger than $T_{\rm min}$.  Since the mass limits are larger than $M_{c}$, these are LBH dual to the QGP with positive heat capacity.  

A remark on the different behaviour of entropy should be emphasized here.  The $m\to 0$ fermionic AdS star at zero temperature at any $\mu$ has vanishing entropy while the radiation star generically has positive entropy for $T>0$.  The gravitational collapse of the AdS star of massless fermions evolves the entropy from zero to the Bekenstein-Hawking entropy~(BH entropy) of the BH at the end of the collapse.  The collapse of the radiation star or thermal AdS space, on the other hand, evolves the entropy from finite positive value to the same BH entropy.  It appears that the $m\to 0$ fermionic AdS star at zero temperature or the ``radiation" star is another thermodynamical phase of the AdS space, a possibly true ground state with zero entropy.  

All the fermionic stars we explored in Section \ref{adsself} are actually in the ground state with zero entropy.  The mass limit ranges from the LBH in the bulk conformal limit to the SBH as $m$ increases.  For suffciently large $m$, the dual QGP to the SBH has negative heat capacity.  Imagine adiabatically injecting fermionic mass into the center of the AdS space until it reaches the mass limit.  Inevitable gravitational collapse of the AdS star into a BH for any $m$ implies that it is possible to form a QGP with negative heat capacity in the dual CFT side.  Once the BH is formed, it starts to radiate at the Hawking temperature.  The Hawking radiation will gradually populate the AdS box but it will not be in thermal equilibrium since the SBH becomes hotter as it radiates.  Conventional argument when the self-gravity of the radiation is neglected is that the SBH with radiation is unstable since the regulated free energy, Eqn.~(\ref{freeE}), is positive.  In a closed system at a fixed temperature~(canonical ensemble), it would radiate or tunnel into the pure radiation.  The gauge dual of the pure radiation in the AdS space is the gas of the conformal cousin of some confined ``hadrons".  The ``hadron" gas is weakly interacting with less number of degrees of freedom comparing to the QGP.  Therefore the typical scenario without self-gravity would be that the QGP with negative heat capacity, once being formed, will eventually radiate~(or condensate) into the ``hadron" gas.  The SBH is also interpreted to be the bounce solution giving tunneling probability of the transition between the thermal AdS and the LBH~\cite{Gross:1982cv,AlvarezGaume:2005fv},
\begin{eqnarray}
\Gamma \propto e^{-B},
\end{eqnarray}
where $B=I_{-}-I_{+}(I_{-})$ for $T<(>) T_{1}$ respectively.  The regulated action $I_{\pm}$ is given by Eqn.~(\ref{IBH}).

At some point, however, the incoming radiation flux becomes equivalent to the outgoing flux and the SBH should settle in a thermal equilibrium with the radiation confined in the AdS space.  On the boundary, the dual picture is the QGP with negative heat capacity gets hotter and hotter after its formation.  Then after the SBH reaches thermal equilibrium with the radiation, the dual QGP would remain at a constant temperature.  It is interesting to obtain a more quantitative picture of the evolution of this QGP with negative heat capacity.  In the following section, we investigate the BH in equilibrium with radiation in the AdS box to see under which condition the back-reaction becomes important.

\subsection{Radiation outside AdS-BH}

One way to find the proportion of radiation and BH energy in a box is to consider the maximal entropy state of the isolated system~\cite{Hawking:1976de}.  In a microcanonical ensemble of BH and radiation with fixed total energy $E = E_{r}+E_{bh}$, the total number of configurations of the system with radiation energy~(entropy) $E_{r}~(S_{r})$ and black hole energy~(entropy) $E_{bh}~(S_{bh})$ is $\exp(S_{r}+S_{bh})$.  The most probable state will be the state which maximizes the total entropy $S_{r}+S_{bh}$.  The general conditions are

\begin{eqnarray}
\frac{\partial S}{\partial E_{r}} = 0, &&   
\frac{\partial^{2}S}{\partial E_{r}^{2}} < 0. \label{cvcon}
\end{eqnarray}
Under the constraint $E=E_{r}+E_{bh}$, the first condition gives $T_{r}=T_{bh}$ since
\begin{equation}
T^{-1} = \frac{\partial S}{\partial E}.  \nonumber
\end{equation}
Using $E_{r}=a T^{d}$, the second condition yields 
\begin{eqnarray}
C_{V}^{-1} & > & -\frac{T_{r}}{d E_{r}},  \nonumber
\end{eqnarray}
where $C_{V}$ is the heat capacity of the BH.  For LBH in the AdS space with positive heat capacity, the condition is trivially satisfied but for SBH, it becomes
\begin{eqnarray}
C_{V} & < & -\frac{E_{r}d}{T_{r}}=-d a T^{d-1}.
\end{eqnarray}
The constant $a$ can be determined when the self-gravity of the radiation is negligible since
\begin{eqnarray}
E_{r} & = & \int_{0}^{\infty} g_{*} T^{1+n} V_{n-1} r^{n-1}~dr, \nonumber \\
   & = & g_{*}V_{n-1}\frac{\sqrt{\pi}}{2}\frac{\Gamma(n/2)}{\Gamma((n+1)/2)}\ell^{n}T^{1+n}= a T^{1+n},  \nonumber
\end{eqnarray}
where $g_{*}$ is given by Eqn.~(\ref{radrp1}).  Substituting $C_{V}, T$ from Eqn.~(\ref{cvbh}), (\ref{hor}), we obtain the condition on the polynomial
\begin{equation}
F(\ell,n)z^{2n-1}-\left( n z^{2}+(n-2)\right)^{n-1}\left( (n-2)-n z^{2} \right) > 0, \label{conx}
\end{equation}
where $z\equiv r_{+}/\ell$ and
\begin{equation}  
F(\ell,n) \equiv \frac{(4\pi)^{n}V_{n-1}(n-1)\ell^{2n-1}}{4G(n+1)a}.
\end{equation}
This is the condition on the size, $z$, of the SBH in equilibrium with the radiation in the AdS box when self-gravity of the radiation is neglected.  If we fix $z$, it can also be served as the condition on the radius $\ell$~(thus effective volume) of the AdS space.  Moreover, we can rewrite the condition in terms of the radiation energy and mass of the BH as
\begin{eqnarray}
E_{r} & < & \frac{M}{n+1}\frac{(nz^{2}+n-2)^{2}}{(1+z^{2})(n-2-nz^{2})}, \label{erbound}
\end{eqnarray}
where $M$ is mass of the BH.  This form is convenient when taking the flat space limit, $\ell\to \infty$, giving
\begin{eqnarray}  
E_{r} & < & M\left(\frac{n-2}{n+1}\right).
\end{eqnarray} 
The condition (\ref{conx}) can also be translated into a condition on the temperature, $T<T_{2}$, where $T_{2}$ is the saturation temperature.  The radiation and the SBH can be in equilibrium only for temperature lower than this saturation temperature.  For $G, \ell =1; g_{b,f}=2$, the inequality can be solved numerically giving $T_{2}=0.315, 0.484$ correspoding to the SBH with size $z_{2} = 0.341, 0.480$ for $n=3, 4$ respectively.  Note that there is a hierachy $T_{2}>T_{\rm min}(=0.276, 0.450$ for $n=3,4$) and $T_{2}\to T_{\rm min}$ as $g_{\rm eff}\to \infty$.

When an SBH in the AdS is in thermal equilibrium with the radiation under the condition~(\ref{cvcon}), the {\it total} heat capacity is still constrained to be {\it negative} and so is the dual QGP.  This can be shown from Eqn.~(\ref{cvcon}) under constraint $E=E_{r}+E_{bh}$,
\begin{eqnarray*}
\frac{\partial^{2} S}{\partial E_{r}^{2}} & = & \frac{\partial^{2} S_{r}}{\partial E_{r}^{2}}+\frac{\partial^{2} S_{bh}}{\partial E_{bh}^{2}}, \\
& = & \frac{\partial}{\partial E_{r}}(\frac{1}{T_{r}})+\frac{\partial}{\partial E_{bh}}(\frac{1}{T_{bh}}), \\
& = & -\frac{1}{T_{r}^{2}}\frac{\partial T_{r}}{\partial E_{r}}-\frac{1}{T_{bh}^{2}}\frac{\partial T_{bh}}{\partial E_{bh}}=-\frac{1}{T^{2}}\left( \frac{C_{V}^{r}+C_{V}^{bh}}{C_{V}^{r} ~C_{V}^{bh}}\right) < 0.
\end{eqnarray*}
Since $C_{V}^{bh}<0$ for the SBH, the total heat capacity must be negative. 

Quantum fluctuations force BH to emit Hawking radiation but after some time, the amount of radiation in the AdS space would saturate the bound given by the condition (\ref{erbound}), i.e. approaching the point where the total heat capacity is zero.  We would expect the same thing to happen to the dual QGP at the boundary.  Once the QGP with negative heat capacity is formed below the temperature $T_{2}$, it will radiate away its energy until the temperature reaches $T_{2}$ saturating the bound given by (\ref{erbound}).  At this configuration, the QGP even with negative heat capacity~(but the total heat capacity of both BH and radiation is zero) remains marginally stable with the surrounding radiation.    

On the contrary, if the formed SBH has corresponding temperature $T>T_{2}$ or is smaller than the critical size $z_{2}$, it will radiate away all of the energy and the AdS space will be in the radiation phase.  In the dual picture, the QGP will undergo a phase transition and become a gas of confined hadron.  Recall that the parameter which determines the size of the BH is the bulk fermion mass $m$, also the quantity signifying the degree of bulk non-conformality.  When $m$ is very small, the mass limit approaches the universal mass limit $M_{\rm limit}(m\to 0)=0.459, 0.913~(n=3,4)$.  The corresponding BH above this mass limit is an LBH.  However, since $M_{\rm limit}$ is a decreasing function with respect to $m$, it will drop below $M_{c}$ for sufficiently large $m$.  Numerical results for $g_{f}=2$ from Fig.~\ref{Mvsm} shows that for $m>0.113, 0.017~(n=3,4)$, $M_{\rm limit}$ corresponds to an SBH with mass less than $M_{c}=0.385, 0.884~(n=3,4)$.  The investigation of the AdS star reveals that not only it is possible to form the QGP with negative heat capacity, it can become stable thermodynamically around $T\lesssim T_{2}$ if the temperature at the formation is smaller than $T_{2}$.  On the other hand, the QGP with negative heat capacity being formed at $T>T_{2}$ will quickly evaporate~(and condensate) completely into the gas of confined hadron.  

The shear viscosity of the positive-$C_{V}$ QGP is given by Ref.~\cite{oai:arXiv.org:hep-th/0405231,Cai:2008in}~(taking the limit $1/b^{2}\to 0$ of the results in Ref.~\cite{Cai:2008in}, since the metric of LBH in global AdS can be well approximated by the planar AdS-BH metric)
\begin{eqnarray}
\eta & = & \frac{A_{n-1}(r_{+})}{16\pi G V}.
\end{eqnarray}
For the planar AdS-BH background, the graviton absorption cross section of the BH is proportional to the correlator of the boundary stress tensor while the shear viscosity is directly related to the correlator in the zero frequency limit.  As a result, the shear viscosity of the dual QGP of the planar AdS-BH and the LBH in AdS is proportional to the horizon area of the BH.  Due to the universality of the equivalence between the low-energy graviton absorption cross section and the horizon area of a BH~\cite{Das:1996we}, it is reasonable to speculate that the shear viscosity of the exotic negative-$C_{V}$ QGP is also proportional to the horizon area of the dual SBH.  If that is the case, the exotic QGP should have smaller shear viscosity~($\eta$) than the positive-$C_{V}$ QGP at the same temperature while preserving the ratio $\eta/s$ since the entropy density $s\equiv S/V$ is also proportional to the horizon area $A_{n-1}$.  

Holographically, if the SBH coexists with radiation in the AdS space, the dual QGP will also coexist with certain amount of bounded quarks in the form of hadron gas.  The mixed phase of QGP soup and hadron gas will gradually change into the QGP with less proportion of hadron gas as we add energy to the system.  At first, temperature will decrease slightly then bounce to increase with the addition of energy after the temperature reaches $T_{\rm min}$.  This picture suggests that the deconfinement phase transition should be continuous and even though the binding potential between quarks and gluons are screened in the QGP phase, there should remain certain bound states in the system.  This is consistent with previous investigations on the meson melting, see e.g. Ref.~\cite{Burikham:2007kp} and references therein.

\section{Conclusions}   \label{VI}

A static holographic star with spherical symmetry will collapse under its own gravity above a certain mass limit.  We study the dependence of the mass limit on the fermion mass of the degenerate fermionic AdS star and the implications to the QGP in the dual picture at the AdS boundary.  The relationship between the fermion mass and the mass limit is numerically established for both $AdS_{4}$ and $AdS_{5}$.  We found that the mass limit is a decreasing function with respect to the fermion mass and the black hole at the mass limit could be both a small and large black hole in the AdS space.  When the fermion mass approaches zero, the mass limit becomes identical to the mass limit of the radiation star due to the bulk conformal symmetry.  In this bulk conformal limit, the mass limit of the AdS star is maximal and universal, suggesting a universal phase transition in the dual gauge picture.  The AdS star with zero fermion mass has zero bulk entropy in contrast to the radiation star which has considerably large entropy.  The multiplicity of the bulk systems corresponding to the same boundary configuration suggests that there might be a multiplicity in the AdS/CFT correspondence mapping, especially when the bulk symmetry is present.  

The fact that gravitational collapse of the AdS star could result in both small and large black hole implies that the QGP formation in the dual gauge picture could result in the QGP with both negative and positive heat capacity.  We thus explore the condition when the QGP with negative heat capacity could be in equilibrium with the radiation using the dual gravity picture.  It is found that a negative-$C_{V}$ QGP with temperature less than a saturation temperature $T_{2}$ could evolve into a negative-$C_{V}$ QGP at $T_{2}$ and become stable thermodynamically without undergoing a confinement phase transition.  On the other hand, the negative-$C_{V}$ QGP with temperature higher than $T_{2}$ will eventually condensate completely into the gas of confined hadron.

The possibility of the coexistence of the SBH and radiation in the AdS space holographically implies the mixed phase of exotic QGP and hadron gas.  It is interesting to experimentally verify if there is a QGP formed in the heavy ion collisions with such properties.  First, it is produced at certain temperature then becomes hotter and eventually saturates at the saturation temperature $T_{2} > T_{crit}\gtrsim T_{\rm min}$.  Addition of energy to the negative-$C_{V}$ QGP will make it colder.  The temperature will continue to drop until it reaches $T_{\rm min}$ then start to raise with the energy.  It is also possible that the exotic QGP could have smaller shear viscosity than the positive-$C_{V}$ QGP at the same temperature with $\eta \propto A_{n-1}(r_{+})=V_{n-1}r_{+}^{n-1}$~(however, the mixture of hadron gas will increase the viscosity).  It will also have smaller entropy density by the same proportion, $s\propto A(r_{+})$, while preserving the ratio $\eta/s$ of the conventional QGP.  Finally, existence of the mixed phase suggests that the deconfinement transition is continuous.  Throughout the transition, the population of the hadron gas will gradually decrease with the temperature.

\section*{Acknowledgments}

P.B. is supported in part by the Thailand Research Fund~(TRF) and Commission on Higher Education~(CHE) under grant RMU5380048.


\begin{thebibliography}{11}

\bibitem{Bekenstein:1973ur} 
  J.~D.~Bekenstein,
  Phys.\ Rev.\ D {\bf 7}, 2333 (1973).

\bibitem{'tHooft:1993gx}
  G.~'t Hooft,
  ``Dimensional reduction in quantum gravity,''  gr-qc/9310026.

\bibitem{Susskind:1994vu}
  L.~Susskind,
  ``The World As A Hologram,''
  J.\ Math.\ Phys.\  {\bf 36}, 6377 (1995)
  [arXiv:hep-th/9409089].

\bibitem{Gubser:1996de} 
  S.~S.~Gubser, I.~R.~Klebanov and A.~W.~Peet,
  Phys.\ Rev.\ D {\bf 54}, 3915 (1996)
  [hep-th/9602135].

\bibitem{Klebanov:1997kc} 
  I.~R.~Klebanov,
  Nucl.\ Phys.\ B {\bf 496}, 231 (1997)
  [hep-th/9702076].

\bibitem{Gubser:1997yh} 
  S.~S.~Gubser, I.~R.~Klebanov and A.~A.~Tseytlin,
  Nucl.\ Phys.\ B {\bf 499}, 217 (1997)
  [hep-th/9703040].

\bibitem{Gubser:1997se} 
  S.~S.~Gubser and I.~R.~Klebanov,
  Phys.\ Lett.\ B {\bf 413}, 41 (1997)
  [hep-th/9708005].

\bibitem{Maldacena:1997re}
  J.~M.~Maldacena,
  ``The large N limit of superconformal field theories and supergravity,''
  Adv.\ Theor.\ Math.\ Phys.\  {\bf 2}, 231 (1998)
  [Int.\ J.\ Theor.\ Phys.\  {\bf 38}, 1113 (1999)]
  [arXiv:hep-th/9711200].

\bibitem{Witten:1998qj}
  E.~Witten,
  ``Anti-de Sitter space and holography,''  Adv.\ Theor.\ Math.\ Phys.\  {\bf 2}, 253 (1998) [hep-th/9802150].

\bibitem{Witten:1998zw} 
  E.~Witten,
  Adv.\ Theor.\ Math.\ Phys.\  {\bf 2}, 505 (1998)
  [hep-th/9803131].

\bibitem{Hawking:1982dh} 
  S.~W.~Hawking and D.~N.~Page,
  Commun.\ Math.\ Phys.\  {\bf 87}, 577 (1983).

\bibitem{deBoer:2009wk} 
  J.~de Boer, K.~Papadodimas and E.~Verlinde,
  JHEP {\bf 1010}, 020 (2010)
  [arXiv:0907.2695 [hep-th]].

\bibitem{Arsiwalla:2010bt}
  X.~Arsiwalla, J.~de Boer, K.~Papadodimas and E.~Verlinde,
  ``Degenerate Stars and Gravitational Collapse in AdS/CFT,''
  JHEP {\bf 1101}, 144 (2011)
  [arXiv:1010.5784 [hep-th]].

\bibitem{dkk}
 U.~H.~Danielsson, E.~Keski-Vakkuri and M.~Kruczenski,
 ``Spherically collapsing matter in AdS, holography, and shellons,"
  Nucl.~Phys.~B {\bf 563}, 279 (1999)
  [arXiv:hep-th/9905227].

\bibitem{ssz}
 E.~Shuryak, Sang-Jin Sin, Ismail Zahed,
  ``A Gravity Dual of RHIC Collisions,"
   J.~Korean~Phys.~Soc. {\bf 50}, 384 (2007)
   [arXiv:hep-th/0511199].

\bibitem{ls}
 S.~Lin and E.~Shuryak,
  ``Toward the AdS/CFT Gravity Dual for High Energy Collisions. 3. Gravitationally Collapsing Shell and Quasiequilibrium,"
  Phys.~Rev.~D {\bf 78}, 125018 (2008)
  [arXiv:0808.0910[hep-th]].

\bibitem{cy}
 P.~M.~Chesler and L.~G.~Yaffe,
  ``Horizon formation and far-from-equilibrium isotropization in supersymmetric Yang-Mills plasma,"
  Phys.~Rev.~Lett. {\bf 102}, 211601 (2009)
  [arXiv:0812.2053[hep-th]].

\bibitem{bm}
 S.~Bhattacharyya and S.~Minwalla,
  ``Weak Field Black Hole Formation in Asymptotically AdS Spacetimes",
  JHEP {\bf 0909}, 034 (2009)
  [arXiv:0904.0464[hep-th]].

\bibitem{bbb}
V.~Balasubramanian, A.~Bernamonti, J.~de~Boer, N.~Copland, B.~Craps, E.~Keski-Vakkuri, B.~Muller, A.~Schafer, M.~Shigemori,
and W.~Staessens, ``Thermalization of Strongly Coupled Field Theories," Phys.~Rev.~Lett. {\bf 106}, 191601 (2011)
[arXiv:1012.4753[hep-th]].

\bibitem{bbbc}
V.~Balasubramanian, A.~Bernamonti, J.~de~Boer, N.~Copland, B.~Craps, E.~Keski-Vakkuri, B.~Muller, A.~Schafer, M.~Shigemori,
and W.~Staessens, ``Holographic Thermalization," Phys.~Rev. D {\bf 84}, 026010 (2011)
[arXiv:1103.2683[hep-th]].

\bibitem{Burikham:2012kn} 
  P.~Burikham and T.~Chullaphan,
  JHEP {\bf 1206}, 021 (2012)
  [arXiv:1203.0883 [hep-th]].

\bibitem{Page:1985em} 
  D.~N.~Page and K.~C.~Phillips,
  Gen.\ Rel.\ Grav.\  {\bf 17}, 1029 (1985).

\bibitem{Vaganov:2007at} 
  V.~Vaganov,
  arXiv:0707.0864 [gr-qc].

\bibitem{Gross:1982cv}
  D.~J.~Gross, M.~J.~Perry and L.~G.~Yaffe,
  ``Instability of Flat Space at Finite Temperature,''  Phys.\ Rev.\ D {\bf 25}, 330 (1982).

\bibitem{AlvarezGaume:2005fv}
  L.~Alvarez-Gaume, C.~Gomez, H.~Liu and S.~Wadia,
  ``Finite temperature effective action, AdS(5) black holes, and 1/N expansion,''  Phys.\ Rev.\ D {\bf 71}, 124023 (2005)
  [hep-th/0502227].

\bibitem{Hawking:1976de} 
  S.~W.~Hawking,
  Phys.\ Rev.\ D {\bf 13}, 191 (1976).

\bibitem{oai:arXiv.org:hep-th/0405231} 
  P.~Kovtun, D.~T.~Son and A.~O.~Starinets,
  Phys.\ Rev.\ Lett.\  {\bf 94}, 111601 (2005)
  [hep-th/0405231].

\bibitem{Cai:2008in} 
  R.~-G.~Cai and Y.~-W.~Sun,
  JHEP {\bf 0809}, 115 (2008)
  [arXiv:0807.2377 [hep-th]].

\bibitem{Das:1996we} 
  S.~R.~Das, G.~W.~Gibbons and S.~D.~Mathur,
  Phys.\ Rev.\ Lett.\  {\bf 78}, 417 (1997)
  [hep-th/9609052].

\bibitem{Burikham:2007kp} 
  P.~Burikham and J.~Li,
  JHEP {\bf 0703}, 067 (2007)
  [hep-ph/0701259 [HEP-PH]].

\end{thebibliography}
\end{document}